\documentclass{article}

\usepackage[english]{babel}
\usepackage{amsmath}
\usepackage{amssymb}
\usepackage{graphicx}
\usepackage{listings}
\usepackage{pdfpages}
\usepackage{xcolor}
\usepackage{ulem}
\usepackage{subcaption}
\usepackage{geometry}
\usepackage{hyperref}
\usepackage{tikz}
\usetikzlibrary{shapes.geometric,calc,shadows.blur,decorations.pathmorphing,fit,decorations.pathreplacing}
\title{Adjusting the range of cell-cell communication enables fine-tuning of cell fate patterns from checkerboard to engulfing}
\author{Simon Schardt and Sabine C. Fischer}
\date{\today}

\begin{document}

\maketitle

\begin{abstract}
    During development, spatio-temporal patterns ranging from checkerboard to engulfing occur with precise proportions of the respective cell fates. Key developmental regulators are intracellular transcriptional interactions and intercellular signaling. We present an analytically tractable mathematical model based on signaling that reliably generates different cell type patterns with specified proportions. Employing statistical mechanics, We derived a cell fate decision model for two cell types. A detailed steady state analysis on the resulting dynamical system yielded necessary conditions to generate spatially heterogeneous patterns. This allows the cell type proportions to be controlled by a single model parameter. Cell-cell communication is realized by local and global signaling mechanisms. These result in different cell type patterns. A nearest neighbor signal yields checkerboard patterns. Increasing the signal dispersion, cell fate clusters and an engulfing pattern can be generated. Altogether, the presented model allows to reliably generate heterogeneous cell type patterns of different kinds as well as desired proportions.
\end{abstract}

\maketitle

\section{Introduction}

Cell fate decisions play an essential role in establishing cellular function during development. In this process, previously indeterminate cells specify themselves into one of several different cell types. In many cases, there is a strong correlation between gene expression patterns and subsequent cell fate. Therefore, it is necessary to understand the dynamics of different genes to unravel the secrets of differentiation.

One prime example of this differentiation process is the differentiation towards neural and epidermal cells in \textit{Drosophila}. Characteristically, epidermal cell progenitors express high levels of transmembrane protein Notch, whereas neural progenitors express low levels of the same \cite{Heitzler1991, Sternberg1993}. A similar example is found in the inner cell mass (ICM) of the preimplantation mouse embryo. There, the transcription factors NANOG and GATA6 have been identified as the earliest markers for the segregation of the ICM into epiblast and primitive endoderm cells, respectively \cite{Mitsui2003, Schrode2014}. Apart from the spatial cell fate distribution, the correct cell fate ratio is also of particular interest \cite{Schroeter2015, Saiz2016, Saiz2020}.

In mathematical models, cell fate decisions are often described by systems of ordinary differential equations (ODE) based on a gene regulatory network (GRN). At the single cell level, toggle switches as models of interactions of two genes have been investigated in great detail \cite{Cherry2000, Huang2007}. These represent mutual inhibition of two proteins combined with auto-activation. As a result, three stable steady states arise with regard to gene expressions that represent the different cell fates. It depends on the initial conditions which state a cell will be attracted to. At the tissue level, experimental studies hint towards the importance of paracrine signals with regards to differentiation \cite{Nichols2009, Yamanaka2010}.

Lateral interaction models have already found their way into the current research landscape. For the Delta-Notch signaling pathway, patterns of alternating cell types have been reconstructed \cite{Collier1996}. For the mouse embryo, models including cell-cell communication due to fibroblast growth factor signaling have been employed to create similar salt-and-pepper/checkerboard patterns \cite{Bessonnard2014, Tosenberger2017}. So far, these studies are concerned with an averaged nearest neighbor signal, i.e. cells do not communicate beyond their nearest neighbor. Further studies suggest that in fact cell fate patterning in the mouse embryo is the result of a complex interplay of cell signaling, cell division, cell sorting and apoptosis \cite{Morris2010, Morris2013, Nissen2017}.

Mathematical modeling allows untangling the individual components and investigating their pattern formation potential. It was previously shown that cell division alone yields cell fate clusters \cite{Liebisch2020}. Simulations of cells sorting due to differential adhesion have been shown to generate engulfing patterns \cite{Revell2019}. This resembles the result of the minimization of the total contact energy \cite{Emily2007}. Here, we focus on the potential of intercellular signaling. In addition to nearest neighbor signaling, we consider signaling that can reach further across a tissue. This builds upon previous ideas for \textit{Drosophila} \cite{DeJoussineau2003,Cohen2010,Chen2014} as well as the mouse embryo \cite{Stanoev2021,Raina2021}.

Based on methods from statistical mechanics \cite{Bintu2005_1, Bintu2005_2, Garcia2011}, we derived a model describing the temporal development of the expressions of two genes. A generalized signal incorporates external influences on cell fate decisions. Performing a detailed stability analysis of the ODE system, we obtained necessary conditions in the form of a parameter interval to always generate a mixture of two different cell types in a tissue. Numerical simulations for an averaged nearest neighbor signal as well as a distance-based signal demonstrate the potential of our model to establish different spatial cell fate patterns ranging from checkerboard via clustering to engulfing patterns. To quantify the different resulting patterns, we employed individualized pair correlation functions (PCFs). A cell type proportion analysis revealed which proportions our model can create, but also which restrictions there are. Our work introduces an easy to control mathematical model for gene expression and our analysis results provide insight into signaling driven pattern formation and cell type proportioning.

\section{Protein interaction model}

\begin{figure}
    \centering
    \begin{tikzpicture}

\newcommand\W{1.2}

\begin{scope}[%
every node/.style={anchor=west, regular polygon, 
regular polygon sides=6,
draw,
minimum width=2.5*\W cm,
outer sep=0,shape border rotate=90,blur shadow={shadow blur steps=5}
},
      transform shape]
\node[fill=white] (I) at (-3.5*\W, -2*\W) {};
\end{scope}

\node (U) at ($(I) + (-0.5*\W, 0)$) {\LARGE $u$};
\node (V) at ($(I) + (0.5*\W, 0)$) {\LARGE $v$};
\node (S) at ($(I) + (0.5*\W, 1.5*\W)$) {\LARGE $s$};

\draw[->] (S) -- (V);
\draw[-|] (S) to [out=180,in=90] (U);
\draw[-|] ($(U)+(0.25*\W,0.1*\W)$) -- ($(V)+(-0.25*\W,0.1*\W)$);
\draw[|-] ($(U)+(0.25*\W,-0.1*\W)$) -- ($(V)+(-0.25*\W,-0.1*\W)$);

\draw[-|] ($(U)+(3*\W,0.3*\W)$) -- ($(V)+(2.8*\W,0.3*\W)$);
\draw[->] ($(U)+(3*\W,-0.3*\W)$) -- ($(V)+(2.8*\W,-0.3*\W)$);


\draw[->,every loop/.style={looseness=1}] (U) edge[in=150,out=210,loop] (U);
\draw[->,every loop/.style={looseness=1}] (V) edge[in=30,out=330,loop] (V);

\node[right=5] (inhib) at ($(V)+(2.8*\W,0.3*\W)$) {Inhibition};
\node[right=5] (activ) at ($(V)+(2.8*\W,-0.3*\W)$) {Activation};

\end{tikzpicture}
    \caption{Illustration of the GRN considered in this study. Inside the cell $u$ and $v$ inhibit each other. In addition to that, they activate themselves. The signal $s$ is an external factor activating $v$.}
    \label{fig: GRN}
\end{figure}
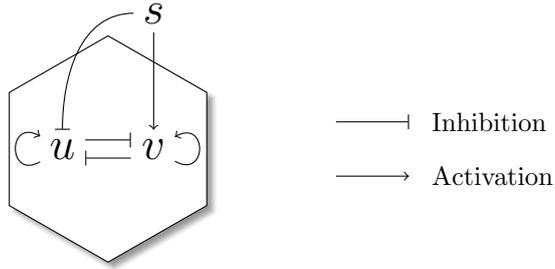

In this section, we derive a model to describe cell fate decisions. As a basis for this we choose methods from \cite{Bintu2005_1, Bintu2005_2, Garcia2011} which allow us to describe transcriptional regulation on the level of the DNA. We consider a simple system of two different transcription factors $u$ and $v$ together with an external signal $s$ describing the cell-cell communication. To this end, we consider a gene regulatory network (GRN) characterized by the mutual inhibition of $u$ and $v$, as well as their auto-activation and the signal $s$ activating $v$ and inhibiting $u$ (Fig. \ref{fig: GRN}).

\subsection{Gene regulation}

To describe the dynamical system underlying transcriptional regulation, we consider two basic assumptions:
\begin{enumerate}
    \item Transcription determines the production of new protein.
    \item Decay describes the lifetime of the protein.
\end{enumerate}
These assumptions are translated into a generic ordinary differential equation (ODE) describing the concentration of a protein $u$ over time:
\begin{equation}
\label{eq: base model}
    \frac{du}{dt} = r_u p_u - \gamma_u u.
\end{equation}
The second term is the exponential decay with decay rate $\gamma_u$. The first term describes the rate of transcription of the corresponding gene. Here, $p_u$ denotes the probability that RNA polymerase (RNAP) is bound to the promoter of $u$. The production rate $r_u$ describes how much protein can be produced while RNAP is bound.

\subsection{Binding probability}

Following \cite{Garcia2011, Bintu2005_1, Bintu2005_2}, we consider the different binding events of a gene regulatory network (GRN). However, we assume that the auto-activatory part of $u$ is dominant, such that the base activity of the RNA polymerase will be neglected. This means that the production of $u$ mainly depends on its binding close to its own promoter. Now the system can be in two different states. Either $u$ is bound or it is not. First we count the number of possibilities how these states might arise. We divide our space into $\Omega$ different lattice sites and describe the total number of protein via $U = u \Omega$. The binomial coefficients yield the number of possible states
\begin{align}
    &\text{Number of unbound states:} &&\frac{\Omega !}{U ! (\Omega - U)!}  \\
    &\text{Number of bound states:} &&\frac{\Omega !}{(U -1)! (\Omega - U + 1)!} 
\end{align}
Assuming different energies whether a protein is unbound $\varepsilon_u^{unbound}$ or bound $\varepsilon_u^{bound}$, the two states have total energies
\begin{align}
    \varepsilon^{unbound} &= U \varepsilon_u^{unbound} \\
    \varepsilon^{bound} &= (U-1) \varepsilon_u^{unbound} + \varepsilon_u^{bound}
\end{align}
Using Boltzmann statistics, the energy of the two states enables us to describe the probability that the system is in either of these states via $e^{-\beta \varepsilon^{unbound}}$ and $e^{-\beta \varepsilon^{bound}}$. The partition function is given by the sum of all possible Boltzmann weights over every microstate, i.e.
\begin{align}
    Z_{total} &= \sum_{\text{microstates}} e^{-\beta\varepsilon_{\text{microstate}}} \\
    &= \frac{\Omega!}{U!(\Omega-U)!}e^{-\beta\varepsilon^{unbound}}  +\frac{\Omega!}{(U-1)!(\Omega-U+1)!}e^{-\beta\varepsilon^{bound}} \\
    &= Z^{unbound} + Z^{bound}
\end{align}
Using the partition function, we are able to calculate the binding probability $p_u$ by the ratio of bound states $Z^{bound}$ and all states combined as $Z^{unbound} + Z^{bound}$
\begin{equation}
\label{eq: binding probability general}
    p_u = \frac{Z^{bound}}{Z^{unbound}+Z^{bound}}
\end{equation}
Assuming $\Omega \gg U$, we use the approximation $\frac{\Omega !}{(\Omega - U)!} \approx \Omega^U$. We divide the numerator and denominator of \eqref{eq: binding probability general} by $Z^{unbound}$ and define the energy difference $\Delta \varepsilon_u := \beta (\varepsilon_u^{bound}-\varepsilon_u^{unbound})$
\begin{equation}
    p_u = \frac{Z^{bound}/Z^{unbound}}{1 + Z^{bound}/Z^{unbound}} = \frac{\frac{U}{\Omega}e^{-\Delta\varepsilon_u}}{1 + \frac{U}{\Omega}e^{-\Delta\varepsilon_u}}.
\end{equation}
For simplicity, we introduce the energy coefficient $\eta_u := e^{-\Delta\varepsilon_u}$ and use $u=U/\Omega$ to get again the volume fractions. This leads to
\begin{equation}
    \label{eq: M-M equation}
    p_u = \frac{\eta_u u}{1 + \eta_u u}.
\end{equation}

\subsection{Interactions}

The crucial parts in transcriptional regulation are the interactions between constituents. In the following, we consider that an additional species $v$ interacts with the promoter of $u$. This results in a system, with the following microstates:
\begin{center}
\begin{tabular}{ | c || c |}
\hline
Binding Event & Number of States \\
\hline\hline
\begin{tabular}{c}
    $U$ unbound \\
    $V$ unbound
  \end{tabular} & $\frac{\Omega !}{U!V!(\Omega !-U-V)!}$ \\
 \hline
\begin{tabular}{c}
    $U$ bound \\
    $V$ unbound
  \end{tabular} & $\frac{\Omega !}{(U-1)!V!(\Omega !-U-V+1)!}$ \\
  \hline
\begin{tabular}{c}
    $U$ unbound \\
    $V$ bound
  \end{tabular} & $\frac{\Omega !}{U!(V-1)!(\Omega !-U-V+1)!}$  \\
 \hline
\begin{tabular}{c}
    $U$ bound \\
    $V$ bound
  \end{tabular} & $\frac{\Omega !}{(U-1)!(V-1)!(\Omega !-U-V+2)!}$ \\
\hline
\end{tabular}
\end{center}
The binding energy differences remain as before with an additional factor for the interaction $\eta_{uv} = e^{-\Delta\varepsilon_{uv}}$. The binding probabilities for $U$ and $V$ are then given by
\begin{equation}
\label{eq: p_u with one interactor}
    p_u = \frac{\eta_u u + \eta_u \eta_v \eta_{uv} uv}{1 +\eta_u u + \eta_v v + \eta_u \eta_v \eta_{uv} uv}
\end{equation}
The advantage or disadvantage given by the interaction energy difference now determines the nature of the interaction. For $\eta_{uv}=1$, \eqref{eq: p_u with one interactor} can be simplified using factorization
\begin{equation}
    p_u = \frac{\eta_u u + \eta_u \eta_v uv}{1 +\eta_u u + \eta_v v + \eta_u \eta_v uv} = \frac{\eta_u u (1 + \eta_v v)}{(1 + \eta_u u)(1 + \eta_v v)} = \frac{\eta_u u}{1 + \eta_u u}.
\end{equation}
The binding probability reduces to the case without interaction. Consequently, the cases where $\eta_{uv} \neq 1$ describe binding probabilities that are either lower or higher than the case with no interaction:
\begin{itemize}
    \item $\eta_{uv} = 0 \quad \Leftrightarrow \quad \Delta\varepsilon_{uv} = \infty$: complete inhibition / blocking
    \item $\eta_{uv} < 1 \quad \Leftrightarrow \quad \Delta\varepsilon_{uv} > 0$: inhibition
    \item $\eta_{uv} = 1 \quad \Leftrightarrow \quad \Delta\varepsilon_{uv} = 0$: no interaction
    \item $\eta_{uv} > 1 \quad \Leftrightarrow \quad \Delta\varepsilon_{uv} < 0$: activation
\end{itemize}
Case $\eta_{uv} = 0$ was listed separately, because it represents a special case of inhibition in which $u$ and $v$ cannot be bound at the same time.

\subsection{Describing the cell fate decision between two fates}

We imagine a system, where two antagonistic proteins $u$ and $v$ are the deciding factors for the decision of a cell's fate. Therefore, $u$ and $v$ mutually inhibit each other. We use a blocking type of inhibition, i.e. the promoter of $u$ is not active as soon as $v$ is bound in the vicinity of $u$'s promoter and vice-versa. We can also interpret this as $u$ not being able to bind if $v$ is already bound. Both $u$ and $v$ are assumed to be dominantly auto-activating, such that the base activity of the RNAP can be neglected. Finally, an external signal $s$ influences $u$ and $v$ in different ways. It activates $v$ by cooperatively binding with $v$. At the same time, $u$ is inhibited by $s$ and the cooperative binding of $u$ and $s$. We assume that for both promoters, the respective energy coefficients are equal. The above considerations lead to interaction coefficients
\begin{equation}
    \label{eq: interaction coefficients}
    \eta_{uv} = \eta_{us} = \eta_{uvs} = 0, \qquad \eta_{vs} \geq 1 \Longleftrightarrow -\Delta\varepsilon_{vs} > 0.
\end{equation}
Any single bound state results in the terms $\eta_\alpha \alpha$ with $\alpha \in \{u,v,s\}$. The remaining state has $v$ and $s$ bound simultaneously, yielding the term $\eta_v\eta_s\eta_{vs}vs$. For the binding probability of $u$, we collect all the terms including $u$ and divide them by the combination of all other terms, resulting in
\begin{equation}
\label{eq: p_u}
    p_u = \frac{\eta_u u}{1 + \eta_v v(1+\eta_s \eta_{vs} s) + \eta_u u + \eta_s s}.
\end{equation}
Likewise, the probability of $v$ is given by
\begin{equation}
\label{eq: p_v}
     p_v = \frac{\eta_v v(1+\eta_s \eta_{vs} s)}{1 + \eta_v v(1+\eta_s \eta_{vs} s) + \eta_u u + \eta_s s}.
\end{equation}
Using \eqref{eq: base model} together with \eqref{eq: p_u} and \eqref{eq: p_v} for a total of $N$ different cells, we end up with a system of ordinary differential equations (ODEs)
\begin{equation}
\label{eq: ODE system}
\begin{aligned}
    \frac{du_i}{dt} &= r_u \frac{\eta_u u_i}{1 + \eta_v v_i(1+\eta_s \eta_{vs} s_i) + \eta_u u_i + \eta_s s_i} - \gamma_u u_i \\
    \frac{dv_i}{dt} &= r_v \frac{\eta_v v_i(1+\eta_s \eta_{vs} s_i)}{1 + \eta_v v_i(1+\eta_s \eta_{vs} s_i) + \eta_u u_i + \eta_s s_i} - \gamma_v v_i, \qquad i = 1,...,N.
\end{aligned}
\end{equation}
We note that so far, the cell-cell interactions are not further specified. This means that the absorbed signals of each cell $s_i$ are provisionally considered as a generalized function of the expression values of all cells
\begin{equation}
    \label{eq: generalized signal}
    \boldsymbol{s}: \mathbb{R}^N \times \mathbb{R}^N \to \mathbb{R}^N: (\boldsymbol{u},\boldsymbol{v}) \mapsto \boldsymbol{s}(\boldsymbol{u},\boldsymbol{v}).
\end{equation}

\section{Steady State Analysis}

\subsection{Existence of steady states}

\begin{figure}[htbp]
\centering
\begin{subfigure}{.32\textwidth}
  \centering
  \includegraphics[width=\linewidth]{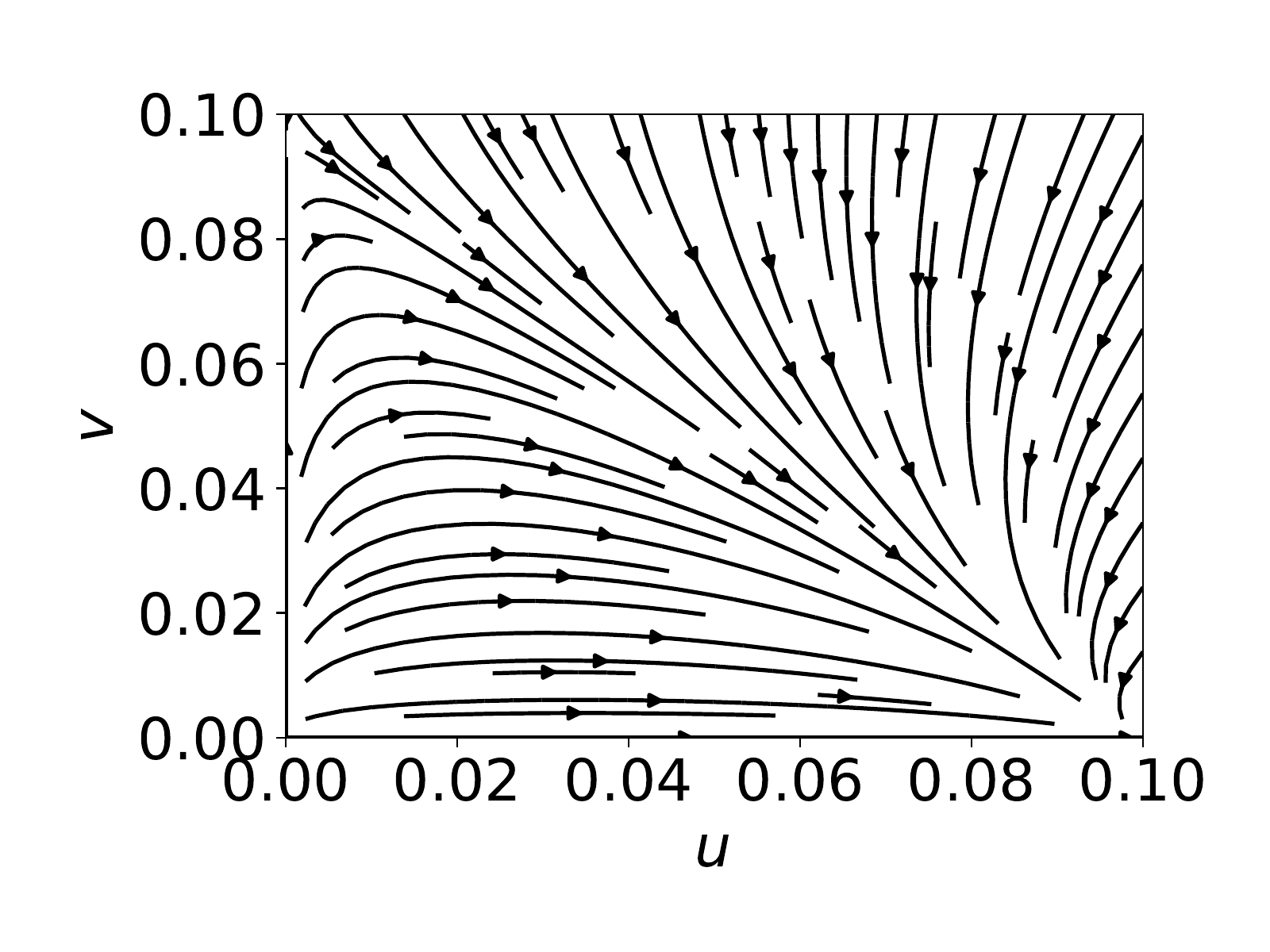}
  \subcaption{$s_i = 0$}
\end{subfigure}
\begin{subfigure}{.32\textwidth}
  \centering
  \includegraphics[width=\linewidth]{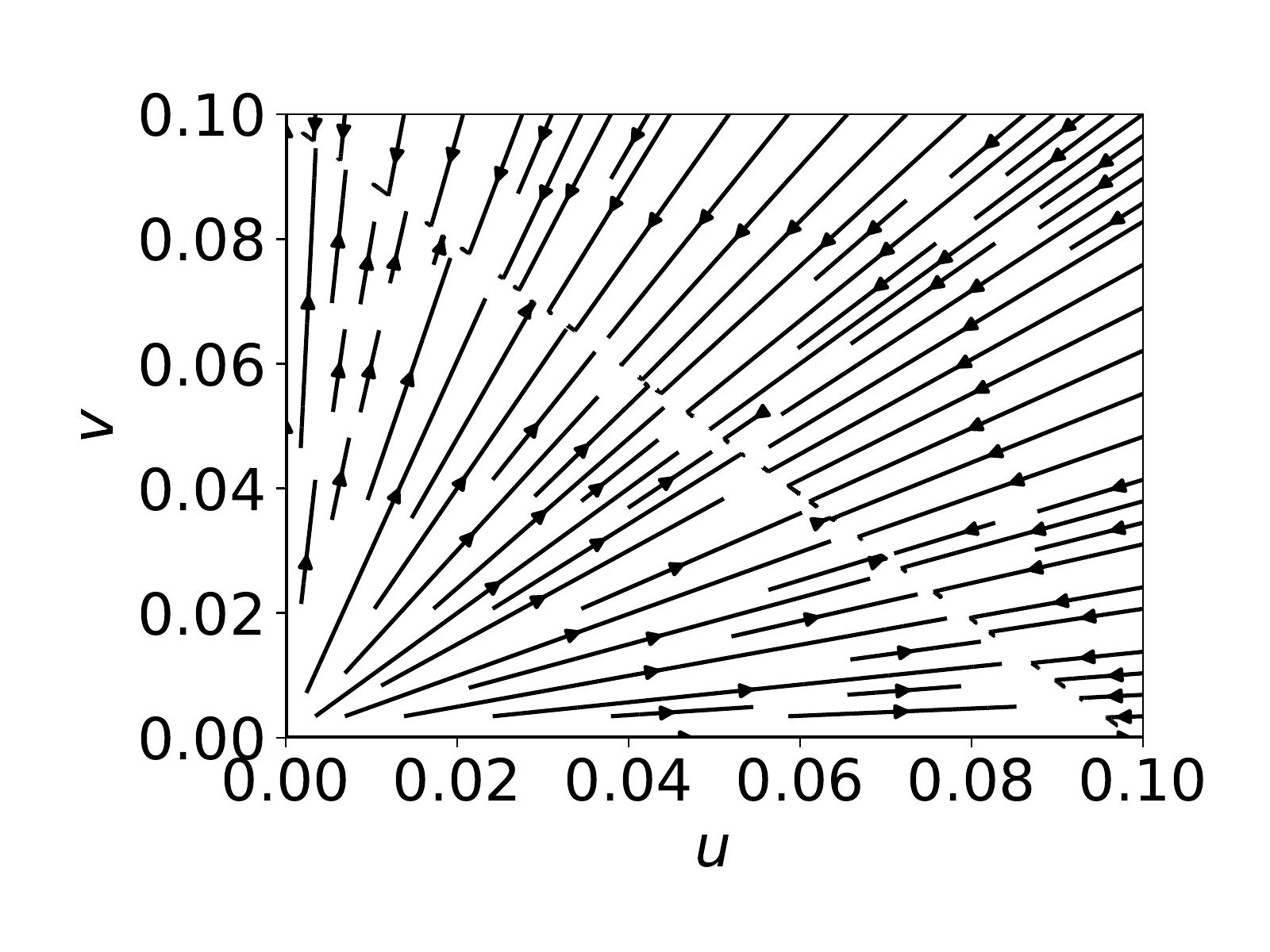}
  \subcaption{$s_i = s^*$}
\end{subfigure}
\begin{subfigure}{.32\textwidth}
  \centering
  \includegraphics[width=\linewidth]{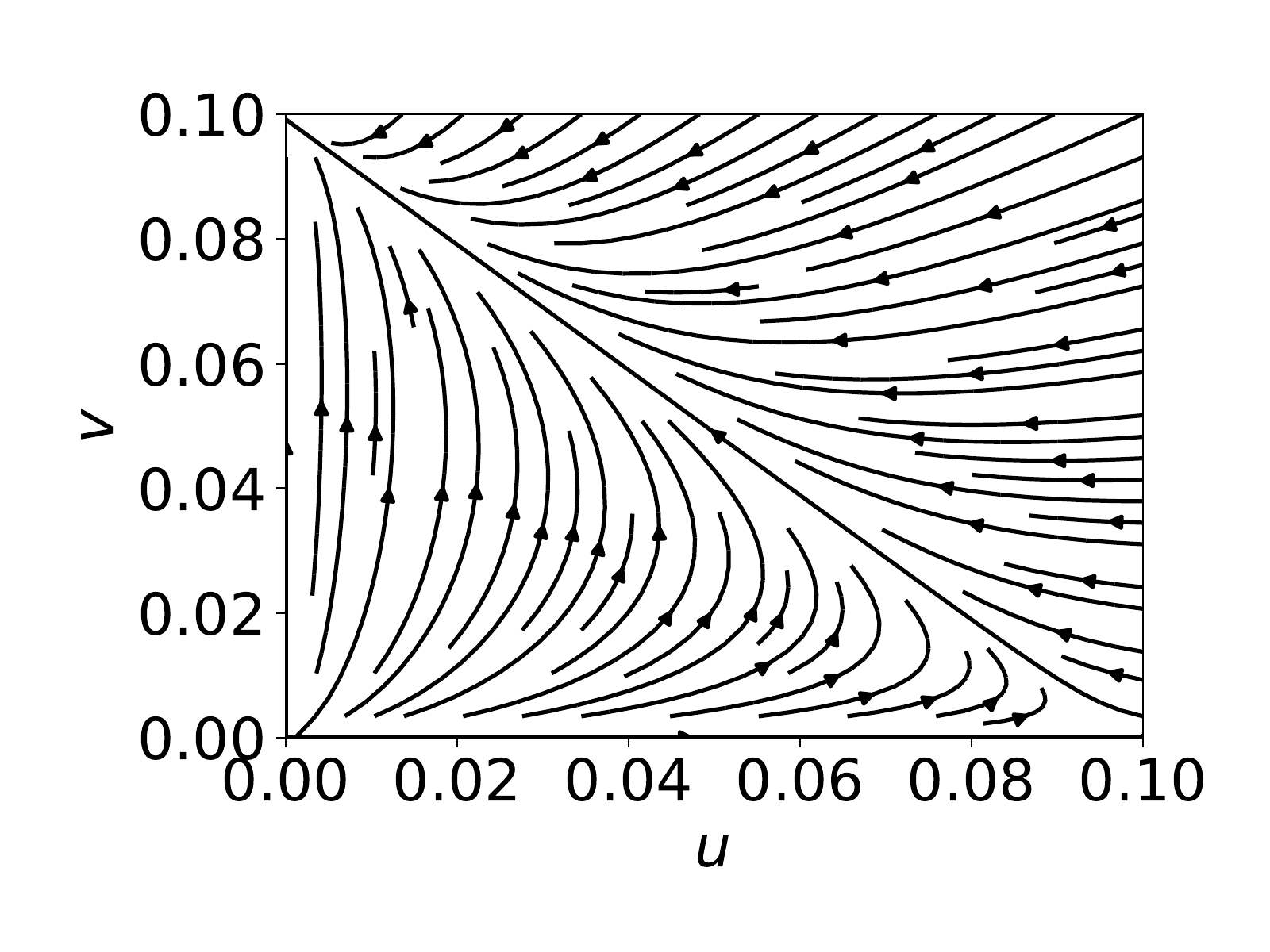}
  \subcaption{$s_i = 2s^*$}
\end{subfigure}
\caption{Streamline phase portraits of ODE system \eqref{eq: ODE system} for a single cell and three different values for $s_i$. Arrows show the path from the initial condition towards the respective steady states \eqref{eq: steady state 2}, \eqref{eq: steady state 4} and \eqref{eq: steady state 3}.}
\label{fig: phase portrait}
\end{figure}

In order to get a better understanding of our ODE system, we want to delve further into the resulting steady states of the system. This means, we consider
$$\frac{du_i}{dt} = 0 = \frac{dv_i}{dt}.$$
Consequently, we get
\begin{align}
\label{eq: steady_state_u}
    \frac{\eta_u u_i}{1 + \eta_u v_i(1+\eta_s\eta_{vs}s_i) + \eta_u u_i + \eta_s s_i} &= \frac{\gamma_u}{r_u} u_i, \\
\label{eq: steady_state_v}
    \frac{\eta_v v_i(1+\eta_s\eta_{vs}s_i)}{1 + \eta_v v_i(1+\eta_s\eta_{vs}s_i) + \eta_u u_i + \eta_s s_i} &= \frac{\gamma_v}{r_v} v_i.
\end{align}
When rearranging \eqref{eq: steady_state_u} and \eqref{eq: steady_state_v}, we find two possible solutions for $u_i$ and $v_i$, respectively. These solutions are
\begin{equation}
    \label{eq: solutions_u_v}
    u_i = \begin{cases}
    0 & \\
    \frac{r_u}{\gamma_u} - \frac{1+\eta_v v_i(1 + \eta_s \eta_{vs} s_i) + \eta_s s_i}{\eta_u}
    \end{cases}, \qquad
    v_i = \begin{cases}
    0 & \\
    \frac{r_v}{\gamma_v} - \frac{1+\eta_u u_i + \eta_s s_i}{\eta_v(1 + \eta_s \eta_{vs} s_i)}
    \end{cases}
\end{equation}
Taking every combination of $u_i$ and $v_i$ from \eqref{eq: solutions_u_v} into account, we end up with four different steady states.
For three of the steady states, we can get either no expression of $u$ and $v$ or high expression of one transcription factor and none for the other:
\begin{alignat}{2}
    \label{eq: steady state 1}
    u_i &= 0, && \quad v_i = 0 \\
    \label{eq: steady state 2}
    u_i &= \frac{r_u}{\gamma_u}-\frac{1+\eta_s s_i}{\eta_u}, && \quad v_i = 0 \\
    \label{eq: steady state 3}
    u_i &= 0, && \quad v_i = \frac{r_v}{\gamma_v}-\frac{1+\eta_s s_i}{\eta_v(1+\eta_s \eta_{vs} s_i)} 
\end{alignat}
These steady states share the lower bound $0$. Additionally, a rough estimate for an upper bound is given by the ratios of reproduction and decay $r_u/\gamma_u$ and $r_v/\gamma_v$. For parameter combinations such that 
\begin{equation}
     \frac{r_u}{\gamma_u} \gg \frac{1}{\eta_u}, \quad \frac{r_v}{\gamma_v} \gg \frac{1}{\eta_v} + \frac{\eta_s}{\eta_v}s_i
\end{equation}
the left hand sides of the inequalities provide a reliable estimate for the steady state values.\\
The fourth steady state is an oddity that arises by combining the non-zero solutions for $u_i$ and $v_i$ from \eqref{eq: solutions_u_v}. When combined, the corresponding variables $u_i$ and $v_i$ cancel out and we find the relation
\begin{equation}
    \label{eq: steady state 4 condition}
    \eta_v (1 + \eta_s \eta_{vs} s_i) = \eta_u \frac{r_u \gamma_v}{r_v \gamma_u}.
\end{equation}
This also leaves our system to be over-determined and the values of $u_i$ and $v_i$ cannot further be identified. However, by using \eqref{eq: steady state 4 condition} in the steady state solution $v_i \neq 0$ in \eqref{eq: solutions_u_v}, we obtain the following state:
\begin{equation}
    \label{eq: steady state 4}
    u_i + \frac{r_u\gamma_v}{r_v\gamma_u} v_i = \frac{r_u}{\gamma_u} - \frac{1+\eta_s s_i}{\eta_u}.
\end{equation}
Isolating $s_i$ in equation \eqref{eq: steady state 4 condition} leads to a critical signal value $s^*$ for which this steady state will always occur
\begin{equation}
    \label{eq: critical signal}
    s^* = \frac{r_u \gamma_v \eta_u - r_v \gamma_u \eta_v}{r_v \gamma_u \eta_v \eta_s \eta_{vs}}.
\end{equation}
This critical signal value is also responsible for a switching behavior in our system (Fig. \ref{fig: phase portrait}). For values below or above $s^*$, a cell ends up in states \eqref{eq: steady state 2} ($u^+v^-$) and \eqref{eq: steady state 3} ($u^-v^+$), respectively. At exactly $s^*$, $u$ and $v$ move towards the straight line defined by \eqref{eq: steady state 4} with no unique steady state. Altogether, we have successfully identified the relevant steady states \eqref{eq: steady state 1}-\eqref{eq: steady state 3} of our ODE system \eqref{eq: ODE system} as well as the condition to force a switch in the cell's fate.

\subsection{Linear stability analysis}

In the following sections, we investigate the steady states in further detail. We employ linear stability analysis to determine the parameter regime that allows us to find a desired steady state for the overall system. At the single cell level, we rule out \eqref{eq: steady state 1}, since it is not relevant to cell fate specification. At the tissue level, we distinguish between homogeneous and heterogeneous steady states. A homogeneous equilibrium state consists of cells of a single type only. This means that either all of the cells in the tissue are in state \eqref{eq: steady state 2} ($u^+v^-$) or all of them are in state \eqref{eq: steady state 3} ($u^-v^+$). Best case scenario, is a mixture of the two cell types. Therefore, we aim at excluding the homogeneous steady states as well. We follow the definition of linear stability for an ODE system
$$\frac{dx_i}{dt} = f(x), \qquad i = 1,...,N.$$
We say, an ODE system is linearly stable in $x^*$, if its linearization matrix $L^{ODE} = f'(x^*)$ has only eigenvalues with negative real part. Using the $N$-dimensional identity matrix $I_N$, we can write the linearization matrix of \eqref{eq: ODE system} as
\begin{equation}
 \label{eq: linearization matrix}
L^{ODE} = \begin{pmatrix}                                
r_u A_{uu} - \gamma_u I_N & r_u A_{uv} \\                
r_v A_{uv} & r_v A_{vv} - \gamma_u I_N                   
\end{pmatrix},
\end{equation}
Using the chain rule, the block matrices $A_{xy}$, $x,y \in \{u,v\}$ can be written in terms of the partial derivatives
\begin{align}
    A_{uu} &= \frac{\partial p_u}{\partial u} + \frac{\partial p_u}{\partial s}\frac{\partial s}{\partial u}  &A_{uv} &= \frac{\partial p_u}{\partial v} + \frac{\partial p_u}{\partial s}\frac{\partial s}{\partial v} \\
    A_{vu} &= \frac{\partial p_v}{\partial u} + \frac{\partial p_v}{\partial s}\frac{\partial s}{\partial u}  &A_{vv} &= \frac{\partial p_v}{\partial v} + \frac{\partial p_v}{\partial s}\frac{\partial s}{\partial v}
\end{align}
where we define $\frac{\partial p_u}{\partial u} := \left(\frac{\partial p_u}{\partial u_j,}(u_i,v_i,s_i)\right)_{i,j=1,...,N}$. The other block matrices are defined analogously. For our purposes, we only need to focus on the following derivatives
\begin{align}
    \label{eq: dpu/du}
    \frac{\partial}{\partial u_j}p_u(u_i,v_i,s_i) &= \begin{cases} \frac{\eta_u(1+\eta_v v_i (1+\eta_s\eta_{vs}s_i) + \eta_s s)}
    {(1 + \eta_v v_i(1+\eta_s \eta_{vs} s_i) + \eta_u u_i + \eta_s s_i)^2}, & \qquad \text{if } i = j \\
    0, & \qquad \text{if } i \neq j
    \end{cases} \\
    \label{eq: dpu/dv}
    \frac{\partial}{\partial v_j}p_u(u_i,v_i,s_i) &= \begin{cases}
    -\frac{\eta_v \eta_u  u_i (1 + \eta_s \eta_{vs}s_i)}
    {(1 + \eta_v v_i(1+\eta_s \eta_{vs} s_i) + \eta_u u_i + \eta_s s_i)^2}, & \qquad \text{if } i = j \\
    0, & \qquad \text{if } i \neq j
    \end{cases} \\
    \label{eq: dpv/du}
    \frac{\partial}{\partial u_j}p_v(u_i,v_i,s_i) &= \begin{cases}
    -\frac{\eta_v \eta_u  v_i (1 + \eta_s \eta_{vs}s_i)}
    {(1 + \eta_v v_i(1+\eta_s \eta_{vs} s_i) + \eta_u u_i + \eta_s s_i)^2}, & \qquad \text{if } i = j \\
    0, & \qquad \text{if } i \neq j
    \end{cases} \\
    \label{eq: dpv/dv}
    \frac{\partial}{\partial v_j}p_v(u_i,v_i,s_i) &= \begin{cases}
    -\frac{\eta_v (1 + \eta_s \eta_{vs}s_i)(1 + \eta_u u_i + \eta_s s_i)}
    {(1 + \eta_v v_i(1+\eta_s \eta_{vs} s_i) + \eta_u u_i + \eta_s s_i)^2}, & \qquad \text{if } i = j \\
    0, & \qquad \text{if } i \neq j
    \end{cases} \\
    \label{eq: dpu/ds}
    \frac{\partial}{\partial s_i}p_u(u_i,v_i,s_i) &= -\frac{\eta_u \eta_s u_i (1 + \eta_v \eta_{vs}v_i)}
    {(1 + \eta_v v_i(1+\eta_s \eta_{vs} s_i) + \eta_u u_i + \eta_s s_i)^2} \\
    \label{eq: dpv/ds}
    \frac{\partial}{\partial s_i}p_v(u_i,v_i,s_i) &= \frac{\eta_v \eta_s v_i (\eta_{vs} + \eta_u \eta_{vs}u_i - 1)}
    {(1 + \eta_v v_i(1+\eta_s \eta_{vs} s_i) + \eta_u u_i + \eta_s s_i)^2}
\end{align}

As usual, the eigenvalues of matrix $L^{ODE}$ are defined as the roots of the characteristic polynomial
\begin{equation}
    \label{eq: characteristic polynomial}
    \chi(\lambda) = \det(L^{ODE}-\lambda I_{2N}),
\end{equation}
where $I_{2N}$ denotes the identity matrix in $2N$ dimensions. At first glance, this determinant seems impossible to calculate. However, when inserting the respective steady states, we are able to reduce the matrix tremendously.

\subsection{Excluding steady state \eqref{eq: steady state 1}}

In the following, we elaborate on how to exclude the first steady state \eqref{eq: steady state 1} as solution for our ODE system \eqref{eq: ODE system}. Without loss of generality, we assume $u_1 = 0 = v_1$. This way, in row $N+1$ all entries but one of the matrix $L^{ODE}-\lambda I_{2N}$ become $0$. The remaining entry with index $(N+1, N+1)$ is
\begin{equation}
    \label{eq: steady state 1 first factor}
    r_v\frac{\partial}{\partial v_1}p_v(0,0,s_1) - \gamma_v - \lambda = r_v\eta_v \frac{1+\eta_s \eta_{vs} s_1}{1+\eta_s s_1} - \gamma_v - \lambda.
\end{equation}
Laplace expansion then enables us to write the determinant of the whole matrix as a product of \eqref{eq: steady state 1 first factor} and the determinant of the remaining submatrix. Thus, it suffices to focus on the first eigenvalue given by
$$r_v\eta_v \frac{1+\eta_s \eta_{vs} s_1}{1+\eta_s s_1} - \gamma_v - \lambda \stackrel{!}{=} 0.$$
This translates to the eigenvalue $\lambda$ being
$$\lambda = r_v\eta_v \frac{1+\eta_s \eta_{vs} s_1}{1+\eta_s s_1} - \gamma_v.$$
Now $\lambda > 0$ yields
$$\eta_v > \frac{\gamma_v}{r_v} \frac{1+\eta_s s_1}{1+\eta_s \eta_{vs} s_1}.$$
Although the signal thus far has not been further specified, we propose a realistic physical representation by assuming $s_i \geq 0$. Furthermore, we consider an activation of $v$ by the signal $s$, i.e. $\eta_{vs}>1$ and therefore, inequality
\begin{equation}
    \label{eq: coefficient restriction 1}
    \eta_v > \frac{\gamma_v}{r_v}
\end{equation}
and consequently
\begin{equation}
    \label{eq: energy restriction 1}
    -\Delta\varepsilon_v > \ln\left(\frac{\gamma_v}{r_v}\right)
\end{equation}
provide the necessary condition for instability. The exclusion of this steady state strengthens our focus on \eqref{eq: steady state 2} and \eqref{eq: steady state 3}, which represent the two different cell types $u^+v^-$ and $u^-v^+$, respectively.

\subsection{Instability of tissue-wide homogeneous steady state \eqref{eq: steady state 2}}

With steady states \eqref{eq: steady state 2} and \eqref{eq: steady state 3}, we aim to find a parameter region for which we achieve a heterogeneous steady state, i.e. we get a tissue with a mixture of cells in the two states. To this end, we derive conditions for instability of the homogeneous steady state. We start with state \eqref{eq: steady state 2} and set $u_i = \frac{r_u}{\gamma_u}-\frac{1+\eta_s s_i}{\eta_u}$ and $v_i = 0$ for all $i$. Inserting these expressions into the derivatives \eqref{eq: dpu/du}-\eqref{eq: dpv/ds} results in a simplification of $L^{ODE}$. Since \eqref{eq: dpv/du} and \eqref{eq: dpv/ds} are zero for every $i,j$, the off-diagonal block matrix $A_{vu}=\boldsymbol{0}$. This means the determinant is given by the product of the determinants of the block matrices on the diagonal. Again, since \eqref{eq: dpv/ds} is zero, $A_{vv}$ becomes a diagonal matrix with diagonal entries
\begin{equation}
    (A_{vv})_i = \frac{\eta_v(1+\eta_s\eta_{vs}s_i)}{1+\eta_u u_i + \eta_s s_i}, \qquad i = 1,...,N
\end{equation}
Inserting $u_i$ yields
\begin{equation}
    (A_{vv})_i = \frac{\gamma_u}{r_u} \frac{\eta_v}{\eta_u} (1+\eta_s\eta_{vs}s_i), \qquad i = 1,...,N.
\end{equation}
Using this, we determine $N$ factors of the characteristic polynomial
\begin{align}
    \chi(\lambda)&= \det\left(r_u A_{uu}-(\gamma_u+\lambda)I_N\right)\det\left(r_v A_{vv}-(\gamma_v+\lambda)I_N\right) \\
    \label{eq: steady state 2 det}
    &= \det\left(r_u A_{uu}-(\gamma_u+\lambda)I_N\right)
    \left[\prod_{i=1}^N \gamma_u\frac{r_v\eta_v}{r_u\eta_u}(1+\eta_s \eta_{vs} s_i) - \gamma_v - \lambda\right]
\end{align}
$N$ eigenvalues are given by the second factor in \eqref{eq: steady state 2 det}. For instability, it is sufficient that only one of these is greater than zero. In other words, this results in the inequality
$$\gamma_u\frac{r_v\eta_v}{r_u\eta_u}(1+\eta_s \eta_{vs} s_i) > \gamma_v.$$
After appropriate rearranging, we obtain a sufficient condition for our parameters
\begin{equation}
\label{eq: coefficient restriction 2}
\eta_u < \eta_v \frac{r_v \gamma_u}{r_u \gamma_v}(1+ \eta_s\eta_{vs} \max_i s_i).
\end{equation}
At this point, the general case cannot be simplified further. Depending on the cell-cell interaction and therefore the incoming signal $s_i$, one can find an even more accurate description of this relation. Alternatively, we can formulate this condition in terms of energy differences as
\begin{equation}
\label{eq: energy restriction 2}
 -\Delta \varepsilon_u < -\Delta \varepsilon_v  + \ln \left(1+ e^{-\Delta\varepsilon_s-\Delta\varepsilon_{vs}} \max_i s_i\right) + \ln \left(\frac{r_v \gamma_v}{r_u \gamma_v}\right),
\end{equation}
which allows us to see the maximum allowed deviation of the difference between $\Delta\varepsilon_u$ and $\Delta\varepsilon_v$. Keep in mind that for this condition, we only relied on the first $N$ eigenvalues. In truth, this condition might be even more relaxed than what we derived.

\subsection{Instability of tissue-wide homogeneous steady state \eqref{eq: steady state 3}}

We set $u_i = 0$ and $v_i = \frac{r_v}{\gamma_v}-\frac{1+\eta_s s_i}{\eta_v(1+\eta_s \eta_{vs} s_i)}$. Using the same approach as before, we find that \eqref{eq: dpu/dv} and \eqref{eq: dpu/ds} are zero for all $i,j$ and thus $A_{uv} = \boldsymbol{0}$. In addition to that, we get a diagonal matrix for $A_{uu}$. For $u_i=0$, its diagonal entries are
\begin{equation}
    (A_{uu})_i = \frac{\eta_u}{1+\eta_v v_i(1+\eta_s\eta_{vs}s_i)+\eta_s s_i}, \qquad i = 1,...,N.
\end{equation}
Inserting $v_i$ yields
\begin{equation}
    (A_{uu})_i = \frac{\gamma_v}{r_v}\frac{\eta_u}{\eta_v}\frac{1}{1+\eta_s\eta_{vs}s_i}, \qquad i = 1,...,N.
\end{equation}
As before, this allows us to determine $N$ factors of the characteristic polynomial
\begin{align}
    \chi(\lambda)&= \det\left(r_u A_{uu}-(\gamma_u+\lambda)I_N\right)\det\left(r_v A_{vv}-(\gamma_v+\lambda)I_N\right) \\
    &= \left[\prod_{i=1}^N \gamma_v\frac{r_u}{r_v}\frac{\eta_u}{\eta_v}\frac{1}{1 + \eta_s \eta_{vs}s_i}-\gamma_u-\lambda\right]\det\left(r_v A_{vv}-(\gamma_v+\lambda)I_N\right).
\end{align}
We exploit again the instability condition that any eigenvalue must be positive and find the inequality
\begin{equation}
    \label{eq: coefficient restriction 3.2}
    \eta_u > \frac{r_v \gamma_u}{r_u \gamma_v}\eta_v(1 + \eta_s \eta_{vs}s_i).
\end{equation}
This yields another condition for $\eta_u$. As before, it is necessary to fulfill this inequality for a single value $s_i$, i.e. the minimum of all possible signal values suffices in that regard
\begin{equation}
    \label{eq: coefficient restriction 3}
    \eta_u > \frac{r_v \gamma_u}{r_u \gamma_v}\eta_v(1 + \eta_s \eta_{vs}\min_i s_i).
\end{equation}
Again, we write this in terms of energy differences
\begin{equation}
    \label{eq: energy restriction 3}
    -\Delta\varepsilon_u  > -\Delta\varepsilon_v + \ln\left(1+ e^{-\Delta\varepsilon_s-\Delta\varepsilon_{vs}} \min_i s_i\right) + \ln \left(\frac{r_v \gamma_u}{r_u \gamma_v}\right).
\end{equation}

\subsection{Steady state summary}

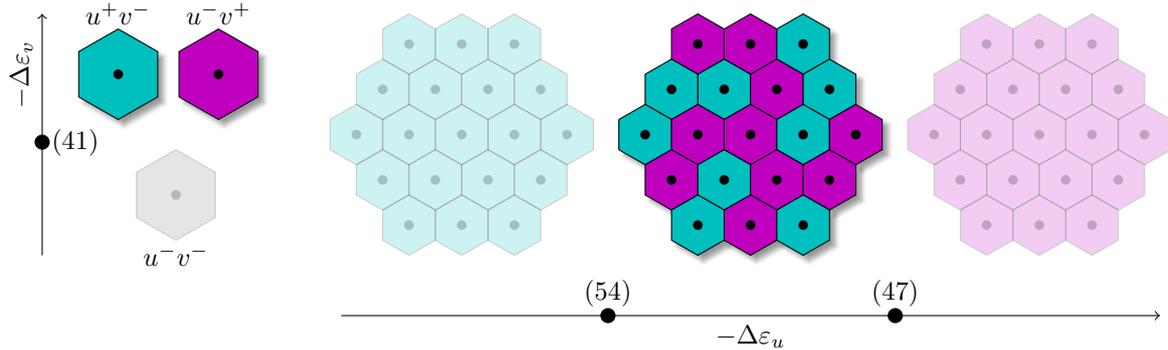
\begin{figure}
    \centering
    \begin{tikzpicture}

\definecolor{NANOG}{rgb}{0.75,0,0.75}
\definecolor{GATA6}{rgb}{0,0.75,0.75}

\newcommand\W{0.8}

\begin{scope}[%
every node/.style={anchor=west, regular polygon, 
regular polygon sides=6,
draw,
minimum width=1.5*\W cm,
outer sep=0,shape border rotate=90
},transform shape]
\node[fill=GATA6,blur shadow={shadow blur steps=5}] (b) at (-5*\W,-0.5*\W) {};
\fill (b) circle (0.08*\W);
\node[fill=gray, opacity=0.2] (a) at ($(b) + (0.3*\W,-2*\W)$) {};
\fill[opacity=0.2, fill opacity=0.2] (a) circle (0.08*\W);
\node[fill=NANOG,blur shadow={shadow blur steps=5}] (c) at ($(b) + (1*\W,0)$) {};
\fill (c) circle (0.08*\W);
\end{scope}

\node[align=left] at ($(a)+(0, -1*\W)$) {$u^-v^-$};
\node[align=left] at ($(b)+(0,1*\W)$) {$u^+v^-$};
\node[align=left] at ($(c)+(0,1*\W)$) {$u^-v^+$};

\fill ($(b) + (-1.25*\W,-1.125*\W)$) circle (0.1) node[right] {\eqref{eq: energy restriction 1}};
\draw[->] ($(b) + (-1.25*\W,-3*\W)$) -- ($(b) + (-1.25*\W,1*\W)$) node[near end, above, rotate=90] {$-\Delta\varepsilon_v$};

\begin{scope}[%
every node/.style={anchor=west, regular polygon, 
regular polygon sides=6,
draw,
minimum width=\W cm,
outer sep=0,shape border rotate=90
},
      transform shape]
    \node[fill=GATA6, opacity=0.2] (A1) {}; \fill[opacity=0.2, fill opacity=0.2] (A1) circle (0.08*\W);
    \node[fill=GATA6, opacity=0.2] (B1) at ($(A1)+(0.433*\W,0)$) {}; \fill[opacity=0.2, fill opacity=0.2] (B1) circle (0.08*\W);
    \node[fill=GATA6, opacity=0.2] (C1) at ($(B1)+(0.433*\W,0)$) {}; \fill[opacity=0.2, fill opacity=0.2] (C1) circle (0.08*\W);
    
    \node[fill=GATA6, opacity=0.2] (A2) at ($(A1)-(2*0.433*\W,3/4*\W)$) {}; \fill[opacity=0.2, fill opacity=0.2] (A2) circle (0.08*\W);
    \node[fill=GATA6, opacity=0.2] (B2) at ($(A2)+(0.433*\W,0)$) {}; \fill[opacity=0.2, fill opacity=0.2] (B2) circle (0.08*\W);
    \node[fill=GATA6, opacity=0.2] (C2) at ($(B2)+(0.433*\W,0)$) {}; \fill[opacity=0.2, fill opacity=0.2] (C2) circle (0.08*\W);
    \node[fill=GATA6, opacity=0.2] (D2) at ($(C2)+(0.433*\W,0)$) {}; \fill[opacity=0.2, fill opacity=0.2] (D2) circle (0.08*\W);
    
    \node[fill=GATA6, opacity=0.2] (A3) at ($(A2)-(2*0.433*\W,3/4*\W)$) {}; \fill[opacity=0.2, fill opacity=0.2] (A3) circle (0.08*\W);
    \node[fill=GATA6, opacity=0.2] (B3) at ($(A3)+(0.433*\W,0)$) {}; \fill[opacity=0.2, fill opacity=0.2] (B3) circle (0.08*\W);
    \node[fill=GATA6, opacity=0.2] (C3) at ($(B3)+(0.433*\W,0)$) {}; \fill[opacity=0.2, fill opacity=0.2] (C3) circle (0.08*\W);
    \node[fill=GATA6, opacity=0.2] (D3) at ($(C3)+(0.433*\W,0)$) {}; \fill[opacity=0.2, fill opacity=0.2] (D3) circle (0.08*\W);
    \node[fill=GATA6, opacity=0.2] (E3) at ($(D3)+(0.433*\W,0)$) {}; \fill[opacity=0.2, fill opacity=0.2] (E3) circle (0.08*\W);
    
    \node[fill=GATA6, opacity=0.2] (A4) at ($(A3)-(0,3/4*\W)$) {}; \fill[opacity=0.2, fill opacity=0.2] (A4) circle (0.08*\W);
    \node[fill=GATA6, opacity=0.2] (B4) at ($(A4)+(0.433*\W,0)$) {}; \fill[opacity=0.2, fill opacity=0.2] (B4) circle (0.08*\W);
    \node[fill=GATA6, opacity=0.2] (C4) at ($(B4)+(0.433*\W,0)$) {}; \fill[opacity=0.2, fill opacity=0.2] (C4) circle (0.08*\W);
    \node[fill=GATA6, opacity=0.2] (D4) at ($(C4)+(0.433*\W,0)$) {}; \fill[opacity=0.2, fill opacity=0.2] (D4) circle (0.08*\W);
    
    \node[fill=GATA6, opacity=0.2] (A5) at ($(A4)-(0,3/4*\W)$) {}; \fill[opacity=0.2, fill opacity=0.2] (A5) circle (0.08*\W);
    \node[fill=GATA6, opacity=0.2] (B5) at ($(A5)+(0.433*\W,0)$) {}; \fill[opacity=0.2, fill opacity=0.2] (B5) circle (0.08*\W);
    \node[fill=GATA6, opacity=0.2] (C5) at ($(B5)+(0.433*\W,0)$) {}; \fill[opacity=0.2, fill opacity=0.2] (C5) circle (0.08*\W);
\end{scope}

\begin{scope}[%
every node/.style={anchor=west, regular polygon, 
regular polygon sides=6,
draw,
minimum width=\W cm,
outer sep=0,shape border rotate=90
},
      transform shape]
    \node[fill=NANOG, opacity=0.2] (A1) at (9.5*\W, 0){}; \fill[opacity=0.2, fill opacity=0.2] (A1) circle (0.08*\W);
    \node[fill=NANOG, opacity=0.2] (B1) at ($(A1)+(0.433*\W,0)$) {}; \fill[opacity=0.2, fill opacity=0.2] (B1) circle (0.08*\W);
    \node[fill=NANOG, opacity=0.2] (C1) at ($(B1)+(0.433*\W,0)$) {}; \fill[opacity=0.2, fill opacity=0.2] (C1) circle (0.08*\W);
    
    \node[fill=NANOG, opacity=0.2] (A2) at ($(A1)-(2*0.433*\W,3/4*\W)$) {}; \fill[opacity=0.2, fill opacity=0.2] (A2) circle (0.08*\W);
    \node[fill=NANOG, opacity=0.2] (B2) at ($(A2)+(0.433*\W,0)$) {}; \fill[opacity=0.2, fill opacity=0.2] (B2) circle (0.08*\W);
    \node[fill=NANOG, opacity=0.2] (C2) at ($(B2)+(0.433*\W,0)$) {}; \fill[opacity=0.2, fill opacity=0.2] (C2) circle (0.08*\W);
    \node[fill=NANOG, opacity=0.2] (D2) at ($(C2)+(0.433*\W,0)$) {}; \fill[opacity=0.2, fill opacity=0.2] (D2) circle (0.08*\W);
    
    \node[fill=NANOG, opacity=0.2] (A3) at ($(A2)-(2*0.433*\W,3/4*\W)$) {}; \fill[opacity=0.2, fill opacity=0.2] (A3) circle (0.08*\W);
    \node[fill=NANOG, opacity=0.2] (B3) at ($(A3)+(0.433*\W,0)$) {}; \fill[opacity=0.2, fill opacity=0.2] (B3) circle (0.08*\W);
    \node[fill=NANOG, opacity=0.2] (C3) at ($(B3)+(0.433*\W,0)$) {}; \fill[opacity=0.2, fill opacity=0.2] (C3) circle (0.08*\W);
    \node[fill=NANOG, opacity=0.2] (D3) at ($(C3)+(0.433*\W,0)$) {}; \fill[opacity=0.2, fill opacity=0.2] (D3) circle (0.08*\W);
    \node[fill=NANOG, opacity=0.2] (E3) at ($(D3)+(0.433*\W,0)$) {}; \fill[opacity=0.2, fill opacity=0.2] (E3) circle (0.08*\W);
    
    \node[fill=NANOG, opacity=0.2] (A4) at ($(A3)-(0,3/4*\W)$) {}; \fill[opacity=0.2, fill opacity=0.2] (A4) circle (0.08*\W);
    \node[fill=NANOG, opacity=0.2] (B4) at ($(A4)+(0.433*\W,0)$) {}; \fill[opacity=0.2, fill opacity=0.2] (B4) circle (0.08*\W);
    \node[fill=NANOG, opacity=0.2] (C4) at ($(B4)+(0.433*\W,0)$) {}; \fill[opacity=0.2, fill opacity=0.2] (C4) circle (0.08*\W);
    \node[fill=NANOG, opacity=0.2] (D4) at ($(C4)+(0.433*\W,0)$) {}; \fill[opacity=0.2, fill opacity=0.2] (D4) circle (0.08*\W);
    
    \node[fill=NANOG, opacity=0.2] (A5) at ($(A4)-(0,3/4*\W)$) {}; \fill[opacity=0.2, fill opacity=0.2] (A5) circle (0.08*\W);
    \node[fill=NANOG, opacity=0.2] (B5) at ($(A5)+(0.433*\W,0)$) {}; \fill[opacity=0.2, fill opacity=0.2] (B5) circle (0.08*\W);
    \node[fill=NANOG, opacity=0.2] (C5) at ($(B5)+(0.433*\W,0)$) {}; \fill[opacity=0.2, fill opacity=0.2] (C5) circle (0.08*\W);
\end{scope}

\begin{scope}[%
every node/.style={anchor=west, regular polygon, 
regular polygon sides=6,
draw,
minimum width=\W cm,
outer sep=0,shape border rotate=90,blur shadow={shadow blur steps=5}
},
      transform shape]
    \node[fill=NANOG] (A1) at (4.75*\W,0) {}; \fill (A1) circle (0.08*\W);
    \node[fill=NANOG] (B1) at ($(A1)+(0.433*\W,0)$) {}; \fill (B1) circle (0.08*\W);
    \node[fill=GATA6] (C1) at ($(B1)+(0.433*\W,0)$) {}; \fill (C1) circle (0.08*\W);
    
    \node[fill=GATA6] (A2) at ($(A1)-(2*0.433*\W,3/4*\W)$) {}; \fill (A2) circle (0.08*\W);
    \node[fill=GATA6] (B2) at ($(A2)+(0.433*\W,0)$) {}; \fill (B2) circle (0.08*\W);
    \node[fill=NANOG] (C2) at ($(B2)+(0.433*\W,0)$) {}; \fill (C2) circle (0.08*\W);
    \node[fill=GATA6] (D2) at ($(C2)+(0.433*\W,0)$) {}; \fill (D2) circle (0.08*\W);
    
    \node[fill=GATA6] (A3) at ($(A2)-(2*0.433*\W,3/4*\W)$) {}; \fill (A3) circle (0.08*\W);
    \node[fill=NANOG] (B3) at ($(A3)+(0.433*\W,0)$) {}; \fill (B3) circle (0.08*\W);
    \node[fill=NANOG] (C3) at ($(B3)+(0.433*\W,0)$) {}; \fill (C3) circle (0.08*\W);
    \node[fill=GATA6] (D3) at ($(C3)+(0.433*\W,0)$) {}; \fill (D3) circle (0.08*\W);
    \node[fill=NANOG] (E3) at ($(D3)+(0.433*\W,0)$) {}; \fill (E3) circle (0.08*\W);
    
    \node[fill=NANOG] (A4) at ($(A3)-(0,3/4*\W)$) {}; \fill (A4) circle (0.08*\W);
    \node[fill=GATA6] (B4) at ($(A4)+(0.433*\W,0)$) {}; \fill (B4) circle (0.08*\W);
    \node[fill=NANOG] (C4) at ($(B4)+(0.433*\W,0)$) {}; \fill (C4) circle (0.08*\W);
    \node[fill=NANOG] (D4) at ($(C4)+(0.433*\W,0)$) {}; \fill (D4) circle (0.08*\W);
    
    \node[fill=GATA6] (A5) at ($(A4)-(0,3/4*\W)$) {}; \fill (A5) circle (0.08*\W);
    \node[fill=NANOG] (B5) at ($(A5)+(0.433*\W,0)$) {}; \fill (B5) circle (0.08*\W);
    \node[fill=GATA6] (C5) at ($(B5)+(0.433*\W,0)$) {}; \fill (C5) circle (0.08*\W);
\end{scope}

\fill ($(A3) - (0.6125*\W,3*\W)$) circle (0.1) node[above] {\eqref{eq: energy restriction 3}};
\fill ($(A3) - (-3.5*\W - 0.6125*\W,3*\W)$) circle (0.1) node[above] {\eqref{eq: energy restriction 2}};
\draw[->] ($(A3) - (5*\W,3*\W)$) -- ($(E3) - (-5*\W,3*\W)$) node[midway, below] {$-\Delta\varepsilon_u$};

\end{tikzpicture}
    \caption{Illustration of the different steady states at the single cell level (left) and the tissue level (right). The states we are aiming for are highlighted with higher opacity. Nodes and their corresponding number on the axes reference the relevant equation for the transition from one state to another.}
    \label{fig: steady states}
\end{figure}

The stability conditions \eqref{eq: energy restriction 2} and \eqref{eq: energy restriction 3} define an interval for $-\Delta\varepsilon_u$,
\begin{equation}
    \label{eq: stability interval}
    \Delta\varepsilon_{min} < -\Delta\varepsilon_u < \Delta\varepsilon_{max}
\end{equation}
with
\begin{align}
    \Delta\varepsilon_{min} &:= -\Delta\varepsilon_v + \ln\left(1+ e^{-\Delta\varepsilon_s-\Delta\varepsilon_{vs}} \min_i s_i\right) + \ln \left(\frac{r_v \gamma_u}{r_u \gamma_v}\right) \\
    \Delta\varepsilon_{max} &:= -\Delta\varepsilon_v + \ln\left(1+ e^{-\Delta\varepsilon_s-\Delta\varepsilon_{vs}} \max_i s_i\right) + \ln \left(\frac{r_v \gamma_u}{r_u \gamma_v}\right)
\end{align}
The reproduction rates $r_u, r_v$ and decay rates $\gamma_u, \gamma_v$ shift this interval by $\ln \left(\frac{r_u \gamma_v}{r_v \gamma_u}\right)$. The length of the interval is determined by the minimum and maximum signal values combined with the associated energy differences $-\Delta\varepsilon_s$ and $-\Delta\varepsilon_{vs}$. The results of our stability analysis are summarized in figure \ref{fig: steady states}. At the single cell level, we are able to exclude $u^-v^-$ cells using inequality \eqref{eq: energy restriction 1}. Therefore, at the tissue level, we can distinguish between three different states. The stability interval \eqref{eq: stability interval} yields the exact parameter regime for the transition of the homogeneous states to the heterogeneous ones. These elegant lower and upper bounds for $-\Delta\varepsilon_u$ incorporate every parameter in our ODE system \eqref{eq: ODE system}. Finally, we know that the lower bound in \eqref{eq: stability interval} is associated with the homogeneous $u^-v^+$ state, whereas the upper bound is associated with the homogeneous $u^+v^-$ state. Therefore, we expect a monotonous increase in the number of $u^+v^-$ cells as the energy difference $-\Delta\varepsilon_u$ increases.

\section{Tissue organization}

\subsection{Cell graph}

\begin{figure}[htbp]
\centering
\begin{subfigure}{.45\textwidth}
  \centering
  \includegraphics[width=\linewidth]{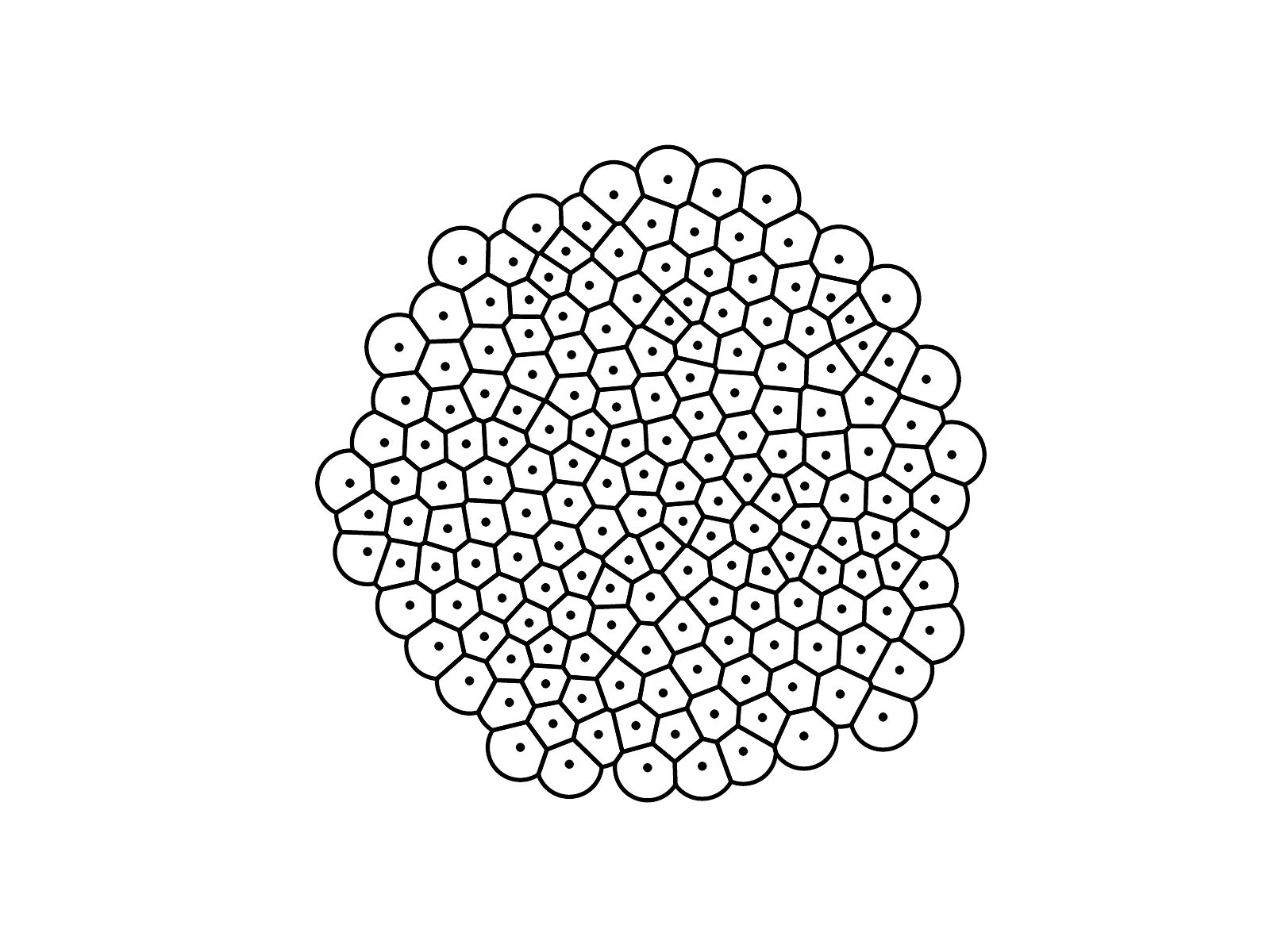}
  \subcaption{Tissue}
\end{subfigure}
\begin{subfigure}{.45\textwidth}
  \centering
  \includegraphics[width=\linewidth]{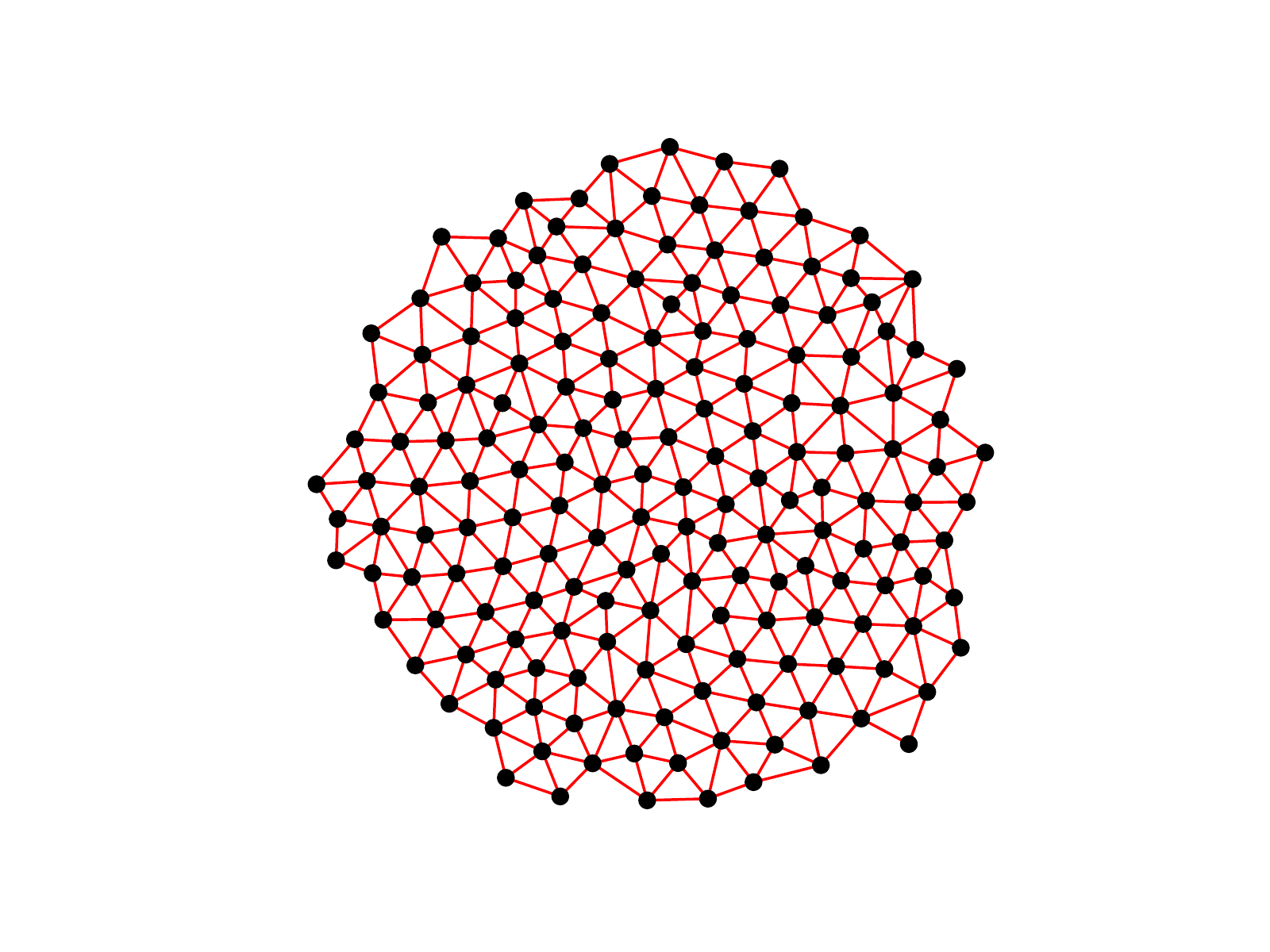}
  \subcaption{Cell graph}
\end{subfigure}
\caption{Visualization of an tissue with $177$ cells (a) and its corresponding cell graph (b). Black lines represent the cell membranes. The cell centroids are shown as black dots in both pictures. Red lines represent the edges, which provide information about which cells are in contact with each other.}
\label{fig: tissue}
\end{figure}

In our context, cells are represented by 2D/3D points in space with a fixed radius which is equal for all cells. The Delaunay cell graph provides a reliable indication of the neighborhood relationships of the cells \cite{Schmitz2017}. Therefore, we initialize our graph $G$ using the Delaunay triangulation. If the Euclidean distance between two cells exceeds the sum of their two radii, then the edge is removed from $G$, i.e. only cells in direct contact with each other are connected via an edge in $G$ (Fig. \ref{fig: tissue}). Edge weights are collectively set to $1$. We then define the cell distance $d_{ij}$ as the length of the shortest path between cells $i$ and $j$.

\subsection{Pair correlation function}

Cell differentiation patterns in our case are the result of two different cell types arising in a tissue. Patterns with the same condition have already been quantified using pair correlation functions (PCFs) \cite{Binder2013}. We use a similar approach to quantify our patterns with a PCF depending on the cell distances $d_{ij}$. This requires counting different types of cell pairings for certain distances. Therefore, we introduce the sets: 
\begin{align}
S_k &= \left\{(i,j)\in \mathbb{N}^2: d_{ij} = k,\, 1 \leq i,j \leq N\right\} \\
S^u_k &= \left\{(i,j)\in S_k: u_i > v_i,\, u_j > v_j\right\} \\
S^v_k &= \left\{(i,j)\in S_k: v_i \geq u_i,\, v_j \geq u_j\right\} \\
T^u &= \left\{i \in \mathbb{N}: u_i > v_i,\, 1 \leq i \leq N\right\} \\
T^v &= \left\{i \in \mathbb{N}: v_i \geq u_i,\, 1 \leq i \leq N\right\}
\end{align}
The pairings of all $u^+v^-$ cells with distance $k$, $S^u_k$, are related to all possible pairings of the same distance $S_k$ by forming their ratios. Analogously, we perform the routine for $u^-v^+$ cell pairings $S^v_k$ to get
\begin{equation}
    r_{uu} = \frac{\vert  S^u_k\vert }{\vert  S_k\vert }, \qquad r_{vv} = \frac{\vert  S^v_k\vert }{\vert  S_k\vert }.
\end{equation}
These ratios alone will not suffice to compare the patterns for varying cell type proportions. Therefore, we normalize these by the probabilities of randomly picking two equal types of cells using the total number of $u^+v^-$ cells $T^u$ and $u^-v^+$ cells $T^v$
\begin{equation}
\label{eq: PCF normalization}
    p_{uu} = \frac{\vert T^u\vert (\vert T^u\vert -1)}{N(N-1)}, \qquad p_{vv} = \frac{\vert T^v\vert (\vert T^v\vert -1)}{N(N-1)}.
\end{equation}
Combined, the PCFs measure the ratios of $u^+v^-$ or $u^-v^+$ cell pairs within every possible distance normalized by the probability of finding these cell pairs, i.e.
\begin{align}
\label{eq: PCF n}
    \rho_u(k) &= \frac{r_{uu}}{p_{uu}} = \frac{\vert S^u_k\vert N(N-1)}{\vert S_k\vert \vert T^u\vert (\vert T^u\vert -1)} \\
\label{eq: PCF g}
    \rho_v(k) &= \frac{r_{vv}}{p_{vv}} = \frac{\vert S^v_k\vert N(N-1)}{\vert S_k\vert \vert T^v\vert (\vert T^v\vert -1)}.
\end{align}
For a uniformly distributed amount of $u^+v^-$ or $u^-v^+$ cells, the correlation function returns a value close to $1$ for every cell distance $k$. Consequently, deviations from $1$ yield information about how much more or fewer equal cell pairs are found in certain ranges.

\section{Numerical results}
In this section, we present the numerical solutions of \eqref{eq: ODE system}. The explicit Euler method is used to solve the ODE until a steady state is reached. We consider two different types of signaling. Paracrine signals that exhibit low diffusivity can be described by a nearest neighbor signal. For larger diffusivities, the signal disperses throughout the tissue such that its intensity decreases with the distance traveled.

\subsection{Nearest neighbor signaling}
\subsubsection{Signal construction}
A signal that is secreted by one cell and diffuses slowly throughout the tissue will likely end up only affecting neighboring cells. We investigate a signal that gets activated by $u$
\begin{equation}
 \label{eq: signal nearest neighbor excluded}
    s_i = \frac{1}{\vert N_G(i)\vert }\sum_{j \in N_G(i)} u_j.
\end{equation}
Here, we used the notation $N_G(i)$ from graph theory to denote the neighbors of vertex $i$ in the graph $G$. We can also write the whole signal in terms of an adjacency matrix. For \eqref{eq: signal nearest neighbor excluded} this matrix will be
\begin{equation}
    \label{eq: adjacency matrix}
    A = (A_{i,j})_{i,j=1,...,M}, \qquad A_{i,j} = \begin{cases}
    \frac{1}{\vert N_G(i)\vert } & \text{if } j \in N_G(i)\\
    0 & \text{if } j \notin N_G(i)
    \end{cases}.
\end{equation}
The signal can ultimately be written as $\boldsymbol{s} = A\boldsymbol{u}$. From the steady state \eqref{eq: steady state 2} we know $u_i = 0$ for some of the cells in a heterogeneous tissue. Therefore, the minimum of the signal will also be $0$. The non zero steady state has a rough upper bound
\begin{equation}
    u_i = \frac{r_u}{\gamma_u} - \frac{1+\eta_s s_i}{\eta_u} < \frac{r_u}{\gamma_u}.
\end{equation}
Therefore, the maximum signal also obeys
\begin{equation}
    \max_i s_i = \max_i \left(\frac{1}{\vert N_G(i)\vert }\sum_{j \in N_G(i)} u_j \right) < \frac{1}{\vert N_G(i)\vert }\sum_{j \in N_G(i)} \frac{r_u}{\gamma_u} = \frac{r_u}{\gamma_u}.
\end{equation}
Using parameter combinations, such that
\begin{equation}
    \frac{r_u}{\gamma_u} \gg \frac{1 + \eta_s \frac{r_u}{\gamma_u}}{\eta_u},
\end{equation}
will transform the upper bound into a proper estimate of the signal values, such that we can conclude
\begin{equation}
    \min_i s_i = 0, \qquad \max_i s_i \approx \frac{r_u}{\gamma_u}.
\end{equation}
Hence, the stability interval can be approximated by
\begin{equation}
    \label{eq: stability interval approx.}
    -\Delta\varepsilon_v + \ln \left(\frac{r_v\gamma_u}{r_u\gamma_v}\right) < -\Delta\varepsilon_u < -\Delta\varepsilon_v + \ln\left(1 + e^{-\Delta\varepsilon_s-\Delta\varepsilon_{vs}} \frac{r_u}{\gamma_u}\right) + \ln \left(\frac{r_v\gamma_u}{r_u\gamma_v}\right).
\end{equation}

\subsubsection{Pattern formation}

\begin{figure}[htbp]
\centering
\begin{subfigure}{.325\textwidth}
  \centering
  \includegraphics[width=\linewidth]{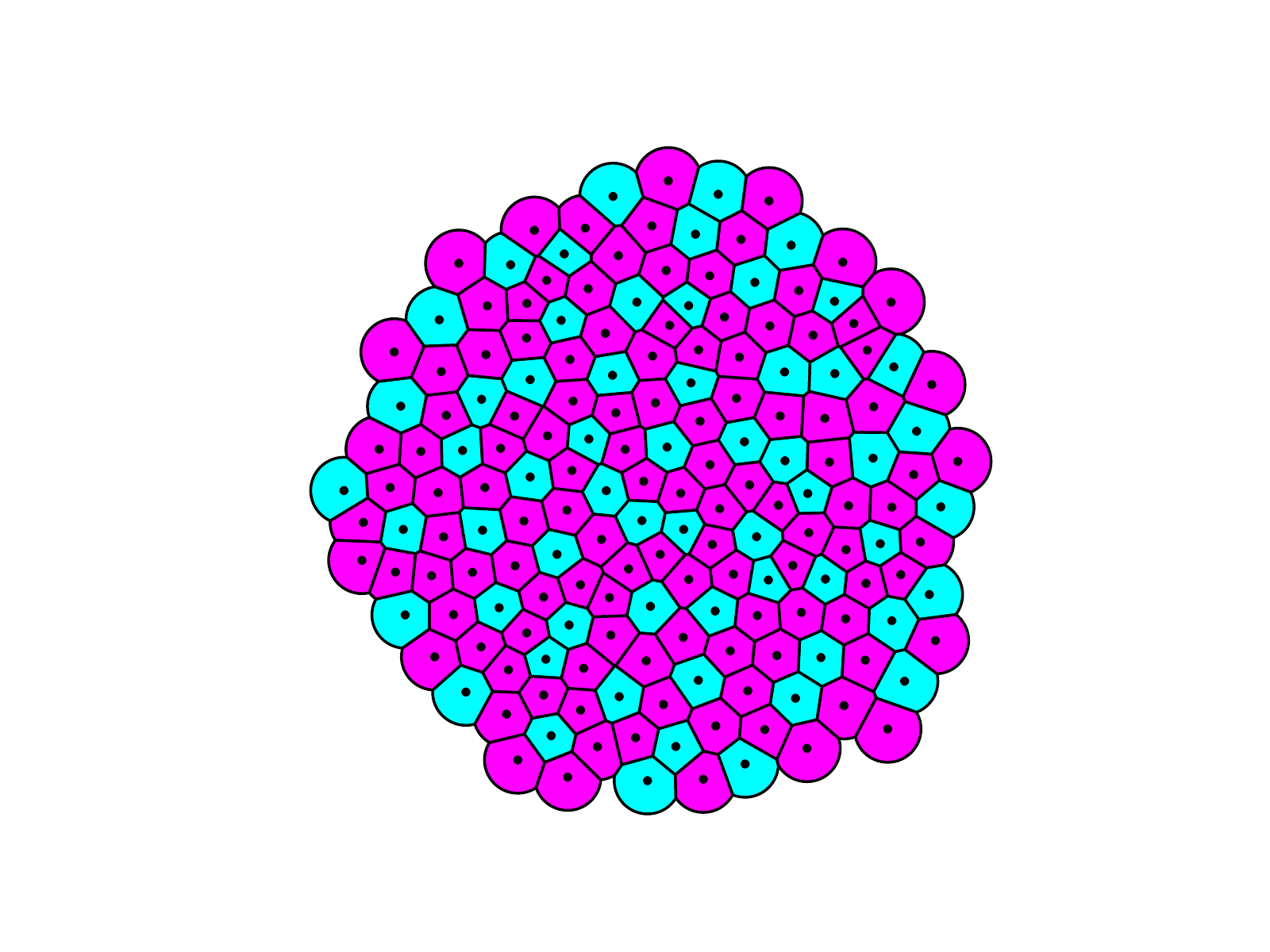}
  \subcaption{$\begin{aligned}
      -\Delta\varepsilon_u &= 7 \\
      \vert T^u\vert :\vert T^v\vert  &= 1:2
  \end{aligned}$}
\end{subfigure}
\begin{subfigure}{.325\textwidth}
  \centering
  \includegraphics[width=\linewidth]{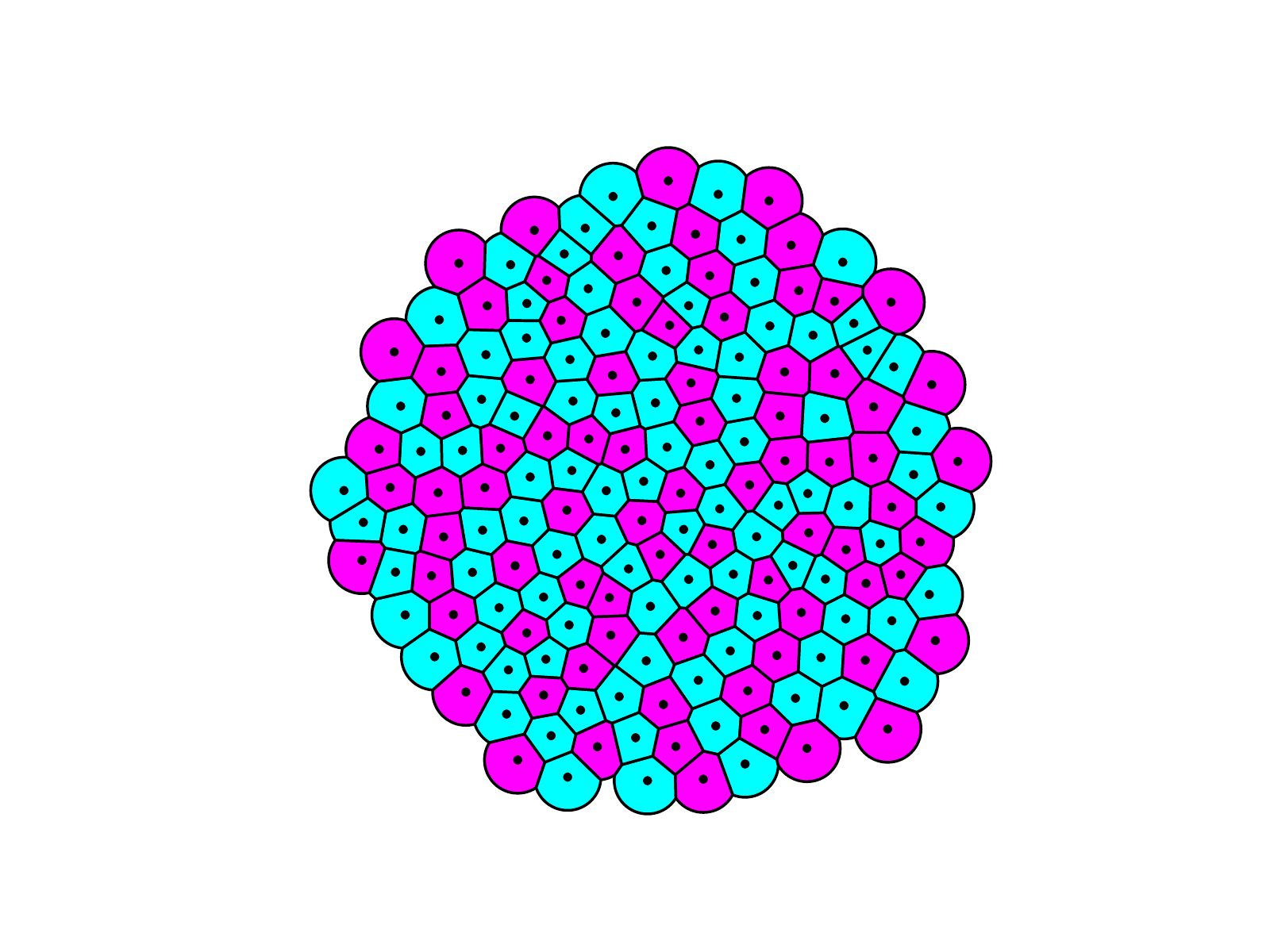}
  \subcaption{$\begin{aligned}
      -\Delta\varepsilon_u &= 7.32 \\
      \vert T^u\vert :\vert T^v\vert  &= 81:96
  \end{aligned}$}
\end{subfigure}
\begin{subfigure}{.325\textwidth}
  \centering
  \includegraphics[width=\linewidth]{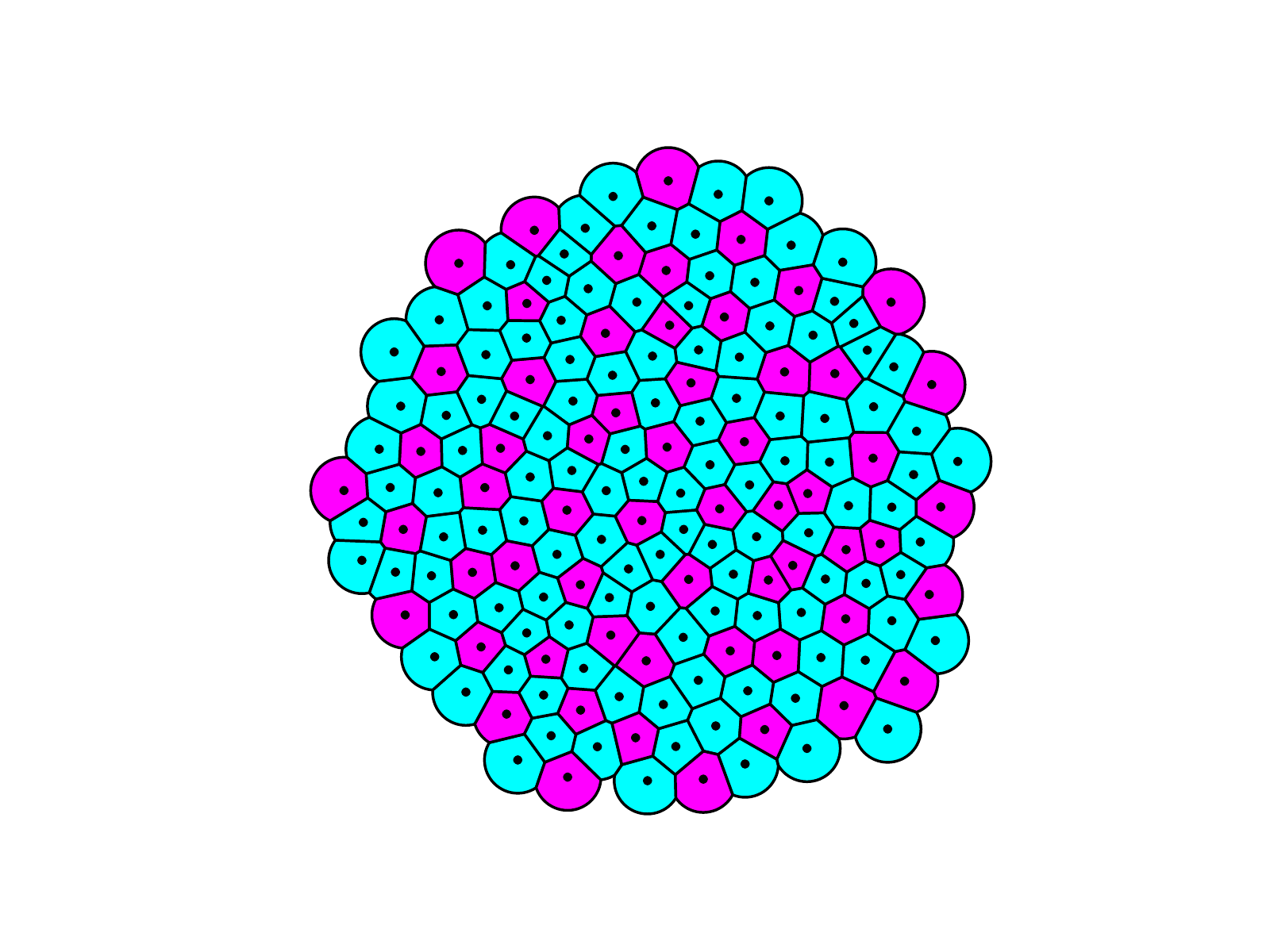}
  \subcaption{$\begin{aligned}
      -\Delta\varepsilon_u &= 7.6 \\
      \vert T^u\vert :\vert T^v\vert  &= 2:1
  \end{aligned}$}
\end{subfigure}
\caption{Checkerboard pattern for three different ratios of $u^+v^-$ and $u^-v^+$ cells. The coloring uses the cell's expression levels for $v$. High $v$ expressions are colored in magenta, low $v$ expressions (high $u$) in cyan.}
\label{fig: checkerboard}
\end{figure}

Models of cell differentiation characterized by lateral inhibition tend to form an approximate checkerboard pattern of cells \cite{Collier1996} with a trend towards alternating cell types wherever possible. In fact, the term "lateral inhibition" comes from the fact that cells of a primary cell fate prevent cells in the environment from adopting the same fate. Despite the name, the interaction between cells in \cite{Collier1996} is caused by an activation rather than inhibition. Our model differs by the inclusion of mutual inhibition and auto-activation. Thus, the goal in this section is to show that our model is still capable of forming checkerboard patterns. The parameter values used in the following simulations are fixed to $-\Delta\varepsilon_v = 6$, $-\Delta\varepsilon_s =-\Delta\varepsilon_{vs} = 2$, $r_u = r_v = 1$ and $\gamma_u = \gamma_v = 10$. The remaining energy difference $-\Delta\varepsilon_u$ is varied based on \eqref{eq: stability interval} to influence the cell type ratio. In the resulting cell fate pattern, $u^+v^-$ cells mostly avoid other $u^+v^-$ cells in their neighborhood (Fig. \ref{fig: checkerboard}). The same behavior is also observed for $u^-v^+$ cells.

\subsubsection{Cell type proportions}

\begin{figure}[htbp]
\centering
\includegraphics[width=0.49\textwidth]{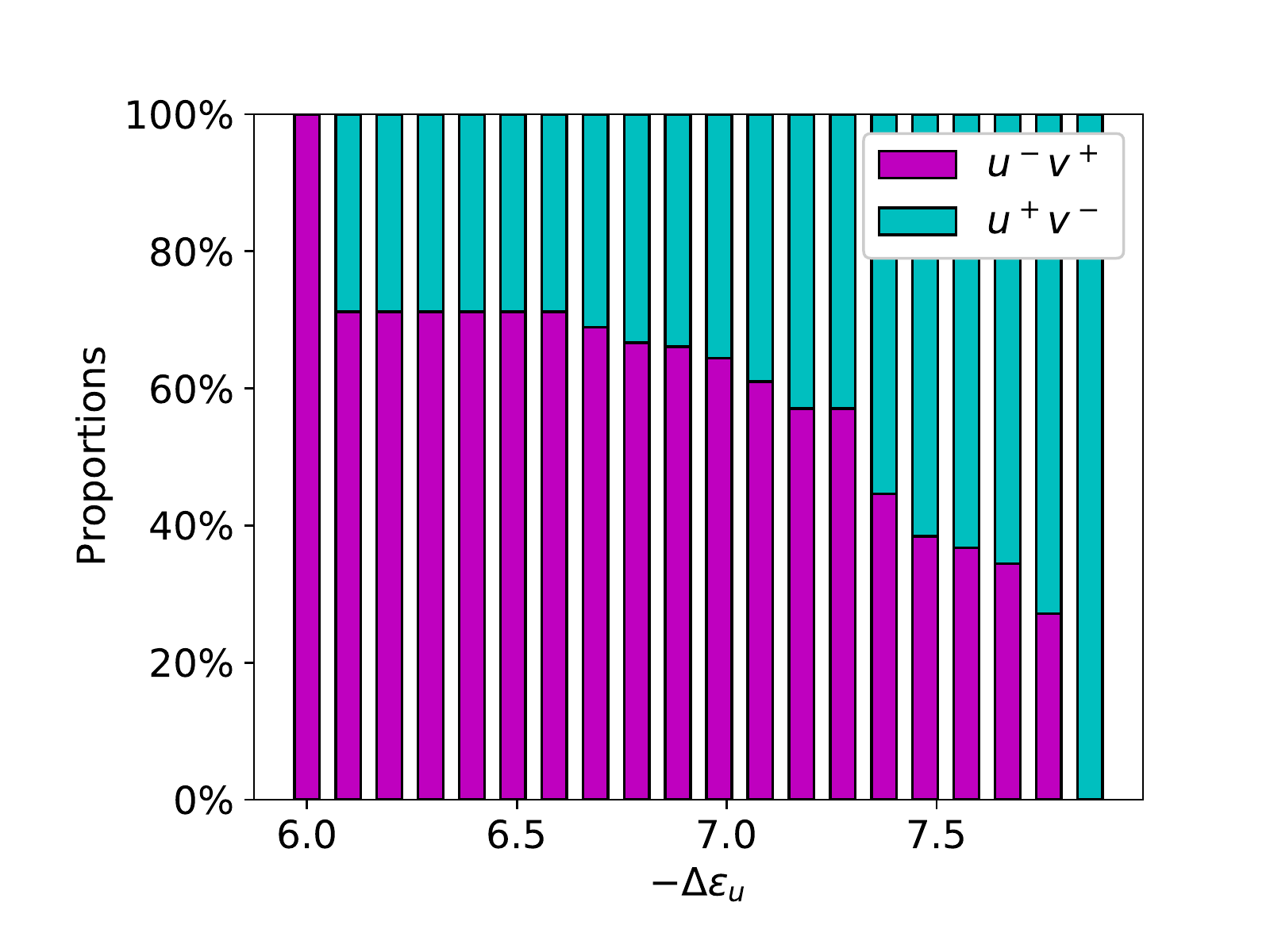}
\caption{Simulated cell type proportions for $20$ equidistant values of $-\Delta\varepsilon_u$ spanning over the stability interval \eqref{eq: stability interval}. Cell proportions for $u^+v^-$ are colored in cyan, $u^-v^+$ in magenta.}
\label{fig: proportions_local}
\end{figure}

In some biological systems, it might be crucial to generate cell types in precise proportions like in the mouse embryo \cite{Saiz2016, Saiz2020}. Hence, we are interested in exploring the capabilities of the model to create certain proportions. The range of possible cell type proportions can be found in the stability interval \eqref{eq: stability interval}. For the parameter combinations chosen in this study, we get 
\begin{equation}
   \frac{1 + \eta_s \frac{r_u}{\gamma_u}}{\eta_u} \leq 0.0043 \ll 0.1 = \frac{r_u}{\gamma_u}.
\end{equation}
Hence, our approximation for the stability interval \eqref{eq: stability interval approx.} is valid and yields the following parameter restrictions for the heterogeneous steady states:
\begin{equation}
\label{eq: bounding interval simulation}
    \eta_u \in (403.43, 2606.08) \qquad \Longleftrightarrow \qquad -\Delta\varepsilon_u \in (6,7.87).
\end{equation}
The various cell type proportions (Fig. \ref{fig: proportions_local}) were simulated by dividing the bounding interval \eqref{eq: bounding interval simulation} into $20$ equidistant values for $-\Delta\varepsilon_u$. The simulation results underline the result of the stability analysis. At the left and right boundaries, we achieve homogeneity. In between, increasing $-\Delta\varepsilon_u$ yields a monotonous transition from only $u^-v^+$ to only $u^+v^-$ cells. The boundary regions suggest that proportions with about $73\%$ of one cell type and $27\%$ of the other are the maximum and minimum cell proportions achievable before reaching homogeneity. An analytical analysis of the relation of the cell type proportions and the parameter $-\Delta\varepsilon_u$ reveals why these jumps occur. Focusing again on a single cell in the tissue, we already identified the tipping point of the cell's fate via equation \eqref{eq: critical signal}. Deviating from $s=s^*$ to $s>s*$ will increase the binding probability for $v$, tipping its fate towards $u^-v^+$. Analogously, $s<s*$ will lead to $u^+v^-$. By definition, $s_i$ is the mean of a cells neighboring $u_j$ values. Assuming the neighbors to be in steady state and using the steady state approximation $u_i \approx r_u/\gamma_u$, the signal can be written as a fraction
\begin{equation}
    s_i = \frac{l_i}{\vert N_G(i)\vert } \frac{r_u}{\gamma_u}, \qquad l_i \in \{0, ..., \vert N_G(i)\vert \},
\end{equation}
where $l_i$ denotes the number of $u^+v^-$ cells adjacent to cell $i$. From this, we can determine the maximum number of $u^+v^-$ cells in a neighborhood for the cell to still adopt the fate $u^+v^-$. Therefore, we replace $s_i$ with $s^*$ and solving the equation for $l_i$ to find
\begin{equation}
    l_i = \vert N_G(i)\vert \frac{\gamma_v}{r_v} \frac{\eta_u - \eta_v}{\eta_v\eta_s \eta_{vs}}.
\end{equation}
Since we are looking for a natural number, the final result is
\begin{equation}
\label{eq: l_max}
    l^{\max} := \left\lfloor \vert N_G(i)\vert \frac{\gamma_v}{r_v} \frac{\eta_u - \eta_v}{\eta_v\eta_s \eta_{vs}}\right\rfloor.
\end{equation}
Here, $\lfloor x \rfloor$ describes the floor function, i.e. the nearest lower integer of a number $x$. Small differences between energy coefficients $\eta_u = \eta_v + \delta$ with $\delta > 0$ being small, will lead to $l^{\max} = 0$. Therefore, a single $u^+v^-$ cell will have no neighbor of equal type. At the same time, cells without any received signal, i.e. $s_i=0$ will adopt $u^+v^-$ fate (Fig. \ref{fig: phase portrait}). In conclusion, a cell surrounded only by $u^-v^+$ cells will adopt $u^+v^-$ fate, whereas a cell with a single $u^+v^-$ in its neighborhood has to adopt $u^-v^+$ fate. On an ideal hexagonal grid, i.e. each cell has exactly six neighbors, an ideal arrangement would amount to $1/3$ of the cells being $u^+v^-$. This estimate nearly fits the simulated proportion jumps of $27\%$ at both ends. An exact number cannot be determined, as the number of neighbors varies from cell to cell with an average of $5.5 \pm 1000000$ neighbors. Further increases of $\eta_u$ only lead to discrete increases of $l^{\max}$, explaining the different jumps in cell type proportions.

\subsection{Distance-based signaling}

\subsubsection{Signal construction}

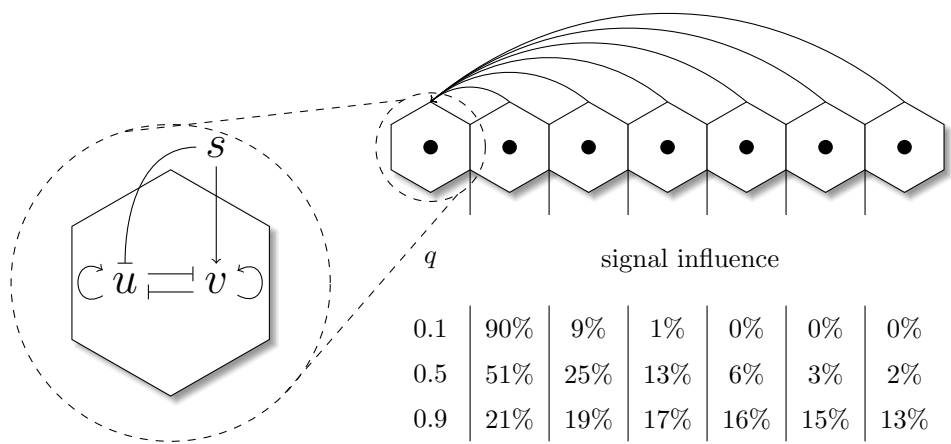
\begin{figure}[htbp]
\centering
\begin{tikzpicture}

\newcommand\W{1.2}

\begin{scope}[%
every node/.style={anchor=west, regular polygon, 
regular polygon sides=6,
draw,
minimum width=\W cm,
outer sep=0,shape border rotate=90,blur shadow={shadow blur steps=5}
},
      transform shape]
    \node[fill=white] (A) at (0*\W,0) {}; \fill (A) circle (0.08*\W);
    \node[fill=white] (B) at ($(A)+(0.433*\W,0)$) {}; \fill (B) circle (0.08*\W);
    \node[fill=white] (C) at ($(B)+(0.433*\W,0)$) {}; \fill (C) circle (0.08*\W);
    \node[fill=white] (D) at ($(C)+(0.433*\W,0)$) {}; \fill (D) circle (0.08*\W);
    \node[fill=white] (E) at ($(D)+(0.433*\W,0)$) {}; \fill (E) circle (0.08*\W);
    \node[fill=white] (F) at ($(E)+(0.433*\W,0)$) {}; \fill (F) circle (0.08*\W);
    \node[fill=white] (G) at ($(F)+(0.433*\W,0)$) {}; \fill (G) circle (0.08*\W);
\end{scope}

\begin{scope}[%
every node/.style={anchor=west, regular polygon, 
regular polygon sides=6,
draw,
minimum width=2.5*\W cm,
outer sep=0,shape border rotate=90,blur shadow={shadow blur steps=5}
},
      transform shape]
\node[fill=white] (I) at (-3.5*\W, -1.5*\W) {};
\end{scope}

\draw [->] ($(B)+(0,0.5*\W)$) to [out=140,in=40] ($(A)+(0,0.5*\W)$);
\draw [->] ($(C)+(0,0.5*\W)$) to [out=140,in=40] ($(A)+(0,0.5*\W)$);
\draw [->] ($(D)+(0,0.5*\W)$) to [out=140,in=40] ($(A)+(0,0.5*\W)$);
\draw [->] ($(E)+(0,0.5*\W)$) to [out=140,in=40] ($(A)+(0,0.5*\W)$);
\draw [->] ($(F)+(0,0.5*\W)$) to [out=140,in=40] ($(A)+(0,0.5*\W)$);
\draw [->] ($(G)+(0,0.5*\W)$) to [out=140,in=40] ($(A)+(0,0.5*\W)$);

\draw[dashed] (A) circle (0.6*\W); 
\draw[dashed] (I) circle (1.75*\W);
\draw[dashed] ($(A) + (-0.49613*0.6*\W, 0.86824*0.6*\W)$) -- ($(I) + (-0.3*1.75*\W, 0.954*1.75*\W)$);
\draw[dashed] ($(A) - (-0.49613*0.6*\W, 0.86824*0.6*\W)$) -- ($(I) - (-0.7*1.75*\W, 0.714*1.75*\W)$);


\node (U) at ($(I) + (-0.5*\W, 0)$) {\LARGE $u$};
\node (V) at ($(I) + (0.5*\W, 0)$) {\LARGE $v$};
\node (S) at ($(I) + (0.5*\W, 1.5*\W)$) {\LARGE $s$};

\draw[->] (S) -- (V);
\draw[-|] (S) to [out=180,in=90] (U);
\draw[-|] ($(U)+(0.25*\W,0.1*\W)$) -- ($(V)+(-0.25*\W,0.1*\W)$);
\draw[|-] ($(U)+(0.25*\W,-0.1*\W)$) -- ($(V)+(-0.25*\W,-0.1*\W)$);

\node () at ($(A)+(0,-1.25*\W)$) {$q$};
\node () at ($(D)+(0.25*\W,-1.25*\W)$) {signal influence};

\draw[->,every loop/.style={looseness=1}] (U) edge[in=150,out=210,loop] (U);
\draw[->,every loop/.style={looseness=1}] (V) edge[in=30,out=330,loop] (V);

\node () at ($(A)+(0,-2*\W)$) {$0.1$};
\node () at ($(B)+(0,-2*\W)$) {$90\%$};
\node () at ($(C)+(0,-2*\W)$) {$9\%$};
\node () at ($(D)+(0,-2*\W)$) {$1\%$};
\node () at ($(E)+(0,-2*\W)$) {$0\%$};
\node () at ($(F)+(0,-2*\W)$) {$0\%$};
\node () at ($(G)+(0,-2*\W)$) {$0\%$};

\node () at ($(A)+(0,-2.5*\W)$) {$0.5$};
\node () at ($(B)+(0,-2.5*\W)$) {$51\%$};
\node () at ($(C)+(0,-2.5*\W)$) {$25\%$};
\node () at ($(D)+(0,-2.5*\W)$) {$13\%$};
\node () at ($(E)+(0,-2.5*\W)$) {$6\%$};
\node () at ($(F)+(0,-2.5*\W)$) {$3\%$};
\node () at ($(G)+(0,-2.5*\W)$) {$2\%$};

\node () at ($(A)+(0,-3*\W)$) {$0.9$};
\node () at ($(B)+(0,-3*\W)$) {$21\%$};
\node () at ($(C)+(0,-3*\W)$) {$19\%$};
\node () at ($(D)+(0,-3*\W)$) {$17\%$};
\node () at ($(E)+(0,-3*\W)$) {$16\%$};
\node () at ($(F)+(0,-3*\W)$) {$15\%$};
\node () at ($(G)+(0,-3*\W)$) {$13\%$};

\draw ($(A)+(0.433*\W,-0.25*\W)$) -- ($(A)+(0.433*\W,-0.75*\W)$);
\draw ($(B)+(0.433*\W,-0.25*\W)$) -- ($(B)+(0.433*\W,-0.75*\W)$);
\draw ($(C)+(0.433*\W,-0.25*\W)$) -- ($(C)+(0.433*\W,-0.75*\W)$);
\draw ($(D)+(0.433*\W,-0.25*\W)$) -- ($(D)+(0.433*\W,-0.75*\W)$);
\draw ($(E)+(0.433*\W,-0.25*\W)$) -- ($(E)+(0.433*\W,-0.75*\W)$);
\draw ($(F)+(0.433*\W,-0.25*\W)$) -- ($(F)+(0.433*\W,-0.75*\W)$);

\draw ($(A)+(0.433*\W,-1.75*\W)$) -- ($(A)+(0.433*\W,-3.25*\W)$);
\draw ($(B)+(0.433*\W,-1.75*\W)$) -- ($(B)+(0.433*\W,-3.25*\W)$);
\draw ($(C)+(0.433*\W,-1.75*\W)$) -- ($(C)+(0.433*\W,-3.25*\W)$);
\draw ($(D)+(0.433*\W,-1.75*\W)$) -- ($(D)+(0.433*\W,-3.25*\W)$);
\draw ($(E)+(0.433*\W,-1.75*\W)$) -- ($(E)+(0.433*\W,-3.25*\W)$);
\draw ($(F)+(0.433*\W,-1.75*\W)$) -- ($(F)+(0.433*\W,-3.25*\W)$);

\end{tikzpicture}
\caption{Illustration of the GRN represented by our model as well as an exemplary representation of the signaling in a one-dimensional cell line. Inside the cell, $u$ and $v$ mutually inhibit each other. Additionally, $v$ gets activated by an extracellular signal, whereas $u$ is inhibited by the same. The signal received by the first cell on the left of the line is the sum of all cell-cell communication between one cell and any other cell in the system. The table highlights how much each cell contributes to the received signal for different dispersions $q \in \{0.1,0.5,0.9\}$. Percentages are rounded to the nearest integer.}
\label{fig: GRN & signal}
\end{figure}

Depending on how a signal disperses in space, not only directly neighboring cells can have an impact on a cell's fate. It is possible, that the collective effect of cells that are further away might also influence its fate decision. Again, the secreted signal of a cell is activated by $u_i$. We define the received signal $s_i$ as the weighted sum of secreted signals over all other cells
\begin{equation}
\label{eq: signal}
    s_i = \left(\sum_{j\neq i}s_j q^{d_{ij}-1}\right)\bigg/\left(\max_k\sum_{j\neq k} q^{d_{kj}-1}\right), \qquad q \in [0,1].
\end{equation}
Here, we use the distances $d_{ij}$ from our cell graph. The weights $q^{d_{ij}-1}$ define the fraction of the signal that gets transported from cell to cell. Let e.g. $q=0.1$, then second nearest neighbors of a cell receive only $10\%$ of the signal of the direct neighbors (Fig. \ref{fig: GRN & signal}). The denominator in \eqref{eq: signal} is used for normalization. It describes the weights of the cell that gets the highest possible signaling weights. In a perfectly arranged circular tissue, this would be the cell right in its center due to the mean of cell distances $d_{ij}$ being lower. The dispersion parameter $q$ enables us to describe the transition from a direct neighbor signal to an equally dispersed signal. For $q=0$, the weights become
\begin{equation}
    q^{d_{ij}-1} = 0^{d_{ij}-1} = \begin{cases}1, \quad \text{for } d_{ij}=1\\0, \quad\text{for } d_{ij}>1
\end{cases}.
\end{equation}
Hence, the weights for all cells that are not directly in contact with the respective cell are $0$ and we obtain a mechanism similar to the local signal \eqref{eq: signal nearest neighbor excluded}. Alternatively, $q=1$ yields
\begin{equation}
    q^{d_{ij}-1} = 1^{d_{ij}-1} = 1.
\end{equation}
This describes the case of every cell having the same impact on other cells independent of the distance between them. In summary, there is a continuous transition from a next neighbor signal at $q=0$, through a distance-based global signal for $q \in [0,1]$, to an evenly distributed signal at $q=1$. In matrix representation we get
\begin{equation}
    \label{eq: dispersion matrix}
    A = (A_{i,j})_{i,j=1,...,M}, \qquad A_{i,j} = \begin{cases}
    a q^{d_{ij}-1} & \text{if } i \neq j\\
    0 & \text{if } i = j
    \end{cases},
\end{equation}
with the normalization factor
\begin{equation}
    a = \left(\max_k\sum_{l\neq k} q^{d_{kl}-1}\right)^{-1}.
\end{equation}
For the estimation of the stability interval, we again use the upper bound $u_i < r_u/\gamma_u$, such that
\begin{align}
    s_i &< \frac{r_u}{\gamma_u}\left(\sum_{j\neq i} q^{d_{ij}-1}\right)\bigg/\left(\max_k\sum_{j\neq k} q^{d_{kj}-1}\right)\\ &\leq \frac{r_u}{\gamma_u}\left(\max_k \sum_{j\neq k} q^{d_{kj}-1}\right)\bigg/\left(\max_k\sum_{j\neq k} q^{d_{kj}-1}\right) = \frac{r_u}{\gamma_u}.
\end{align}
At this point, we realize that the estimation follows the exact same procedure as before, leading to \eqref{eq: stability interval approx.}.

\subsubsection{Pattern formation}
\begin{figure}[htbp]
\centering
\input{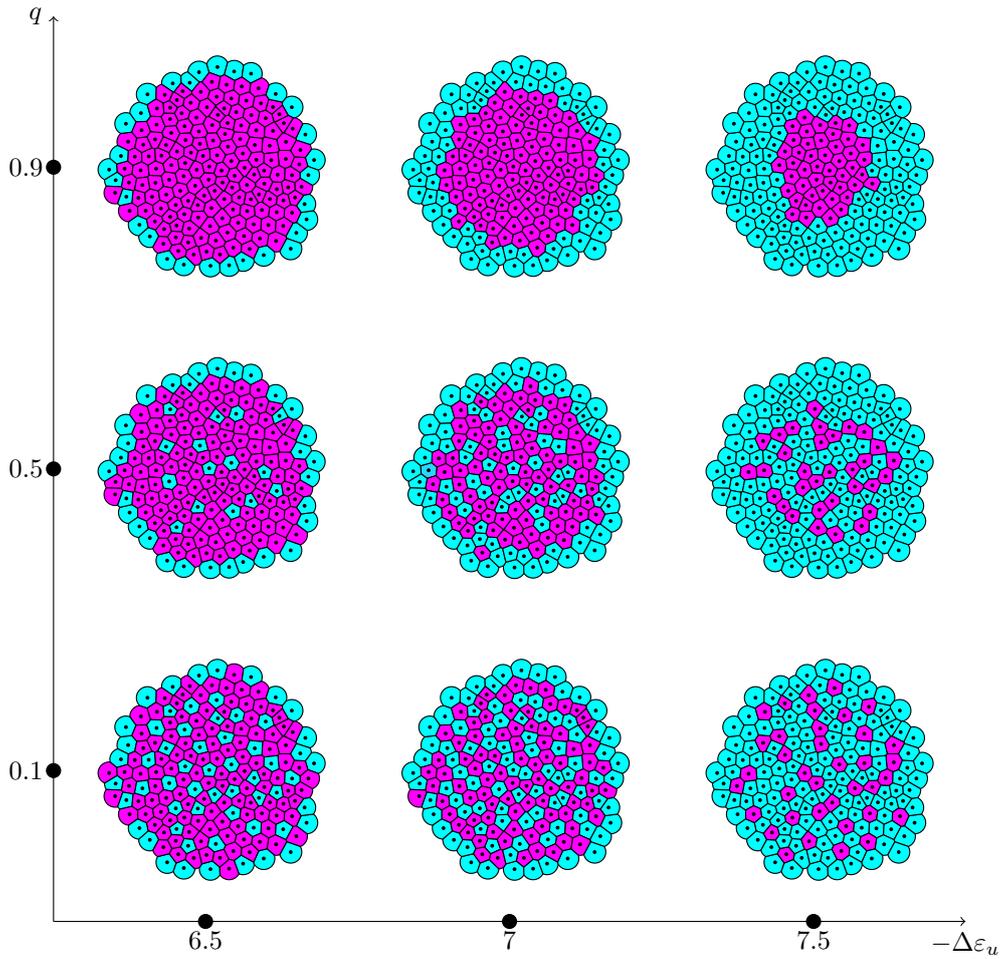}
\caption{Different patterns generated by the model on a tissue geometry with $177$ cells. Colors depict the values of $v_i$ in steady state. High values of $v_i$ correspond to low values in $u_i$ and vice-versa, i.e. cyan and magenta represent $u^+v^-$ and $u^-v^+$ cells, respectively. From left to right, $-\Delta\varepsilon_u$ increases. From bottom to top, the dispersion $q$ increases.}
\label{fig: patterns global}
\end{figure}

\begin{figure}[htbp]
\centering
\begin{subfigure}{.49\textwidth}
  \centering
  \includegraphics[width=\linewidth]{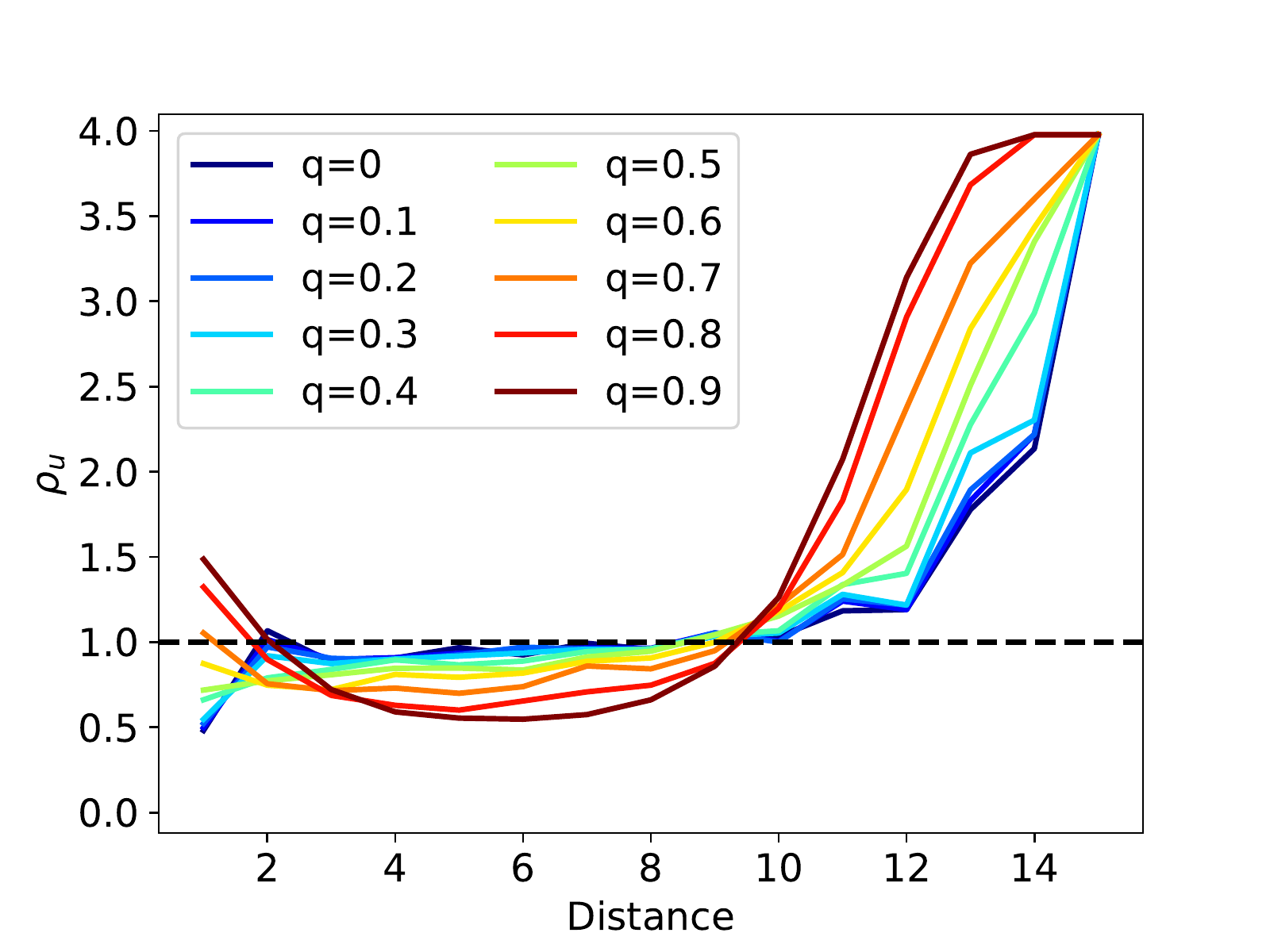}
\end{subfigure}
\begin{subfigure}{.49\textwidth}
  \centering
  \includegraphics[width=\linewidth]{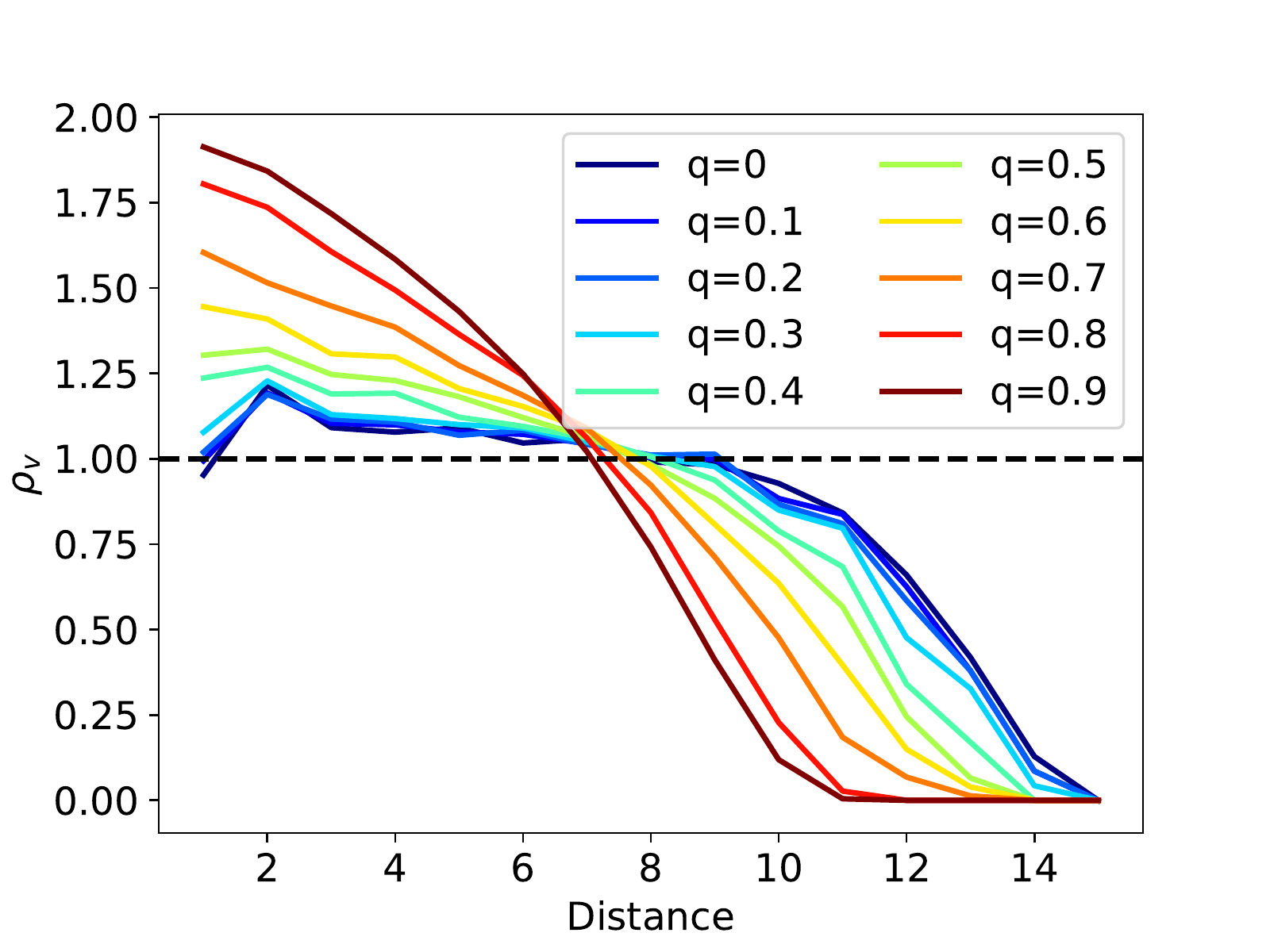}
\end{subfigure}
\caption{PCFs for $u^+v^-$ cells (left) and $u^-v^+$ cells (right) for different dispersion parameters $q$. Any PCF represents a tissue with a ratio of $u^+v^-$:$u^-v^+= 88:89$. The dashed black line at $1$ resembles the PCF values of an ideal uniform distribution of two different cell types. If values lie above $1$, this means there are more pairs found at that distance. Consequently, values below $1$ resemble fewer pairs.}
\label{fig: pair correlations non-local}
\end{figure}

We want to investigate the effect of the distance-based signal on the formation of the patterns. Therefore, we showcase nine simulation results of organoids with different cell type proportions and different signal dispersions (Fig \ref{fig: patterns global}). The patterns generated for $q=0.1$ can mostly be considered of the checkerboard type. In contrast to the averaged nearest neighbor signal, the signal in this case is not averaged over the number of neighbors. Cells at the boundary typically have three to four neighboring cells, whereas cells in the bulk area have a mean of six neighbors. Therefore, cells at the boundary will potentially not be able to get the same amount of signal as cells in the bulk area. The received signal however, is the deciding factor with regard to the cell fate decision in our model. The low amounts of signal received at the boundary make them more likely to adopt the $u^+v^-$ fate.\\
As $q$ increases, we see a higher accumulation of $u^+v^-$ cells near the boundary with a slight clustering behavior in the bulk. For $q=0.9$, the signal disperses strongly enough to generate an engulfing pattern, where $u^-v^+$ cells are completely surrounded by $u^+v^-$ cells. \\
The pattern formation with respect to $q$ can be quantified using the PCFs for both $u^+v^-$ and $u^-v^+$ cells (Fig. \ref{fig: pair correlations non-local}). For comparison, we used a bisection on the stability interval to find values for $-\Delta\varepsilon_u$ that lead to a ratio of $89:88$ $u^+v^-$ and $u^-v^+$ cells for every single $q$. We discover that an increase in $q$ leads to a decrease of $\rho_v$ for large distances, i.e. less and less pairs of $u^-v^+$ cells pairing in the boundary regions. Simultaneously, it increases for small distances due to the cells accumulating in the center. For $\rho_u$, we see a slight increase for large $q$ for small distances and a tremendous one for large distances for all $q$. The slight increase at small distances comes from the fact that the $u^+v^-$ cells arrange in layers at the boundary. The values for intermediate distances slightly decrease as the corresponding regions become more and more devoid of $u^-v^+$ pairs. In conclusion, a distance-based signal according to \eqref{eq: signal} generates patterns ranging from checkerboard to engulfing by increasing the dispersion parameter $q$. Additionally, the PCFs capture the characteristics of these patterns, making it a powerful tool for pattern identification and comparison.

\subsubsection{Cell type proportion}

\begin{figure}[htbp]
\centering
\begin{subfigure}{.32\textwidth}
  \centering
  \includegraphics[width=\linewidth]{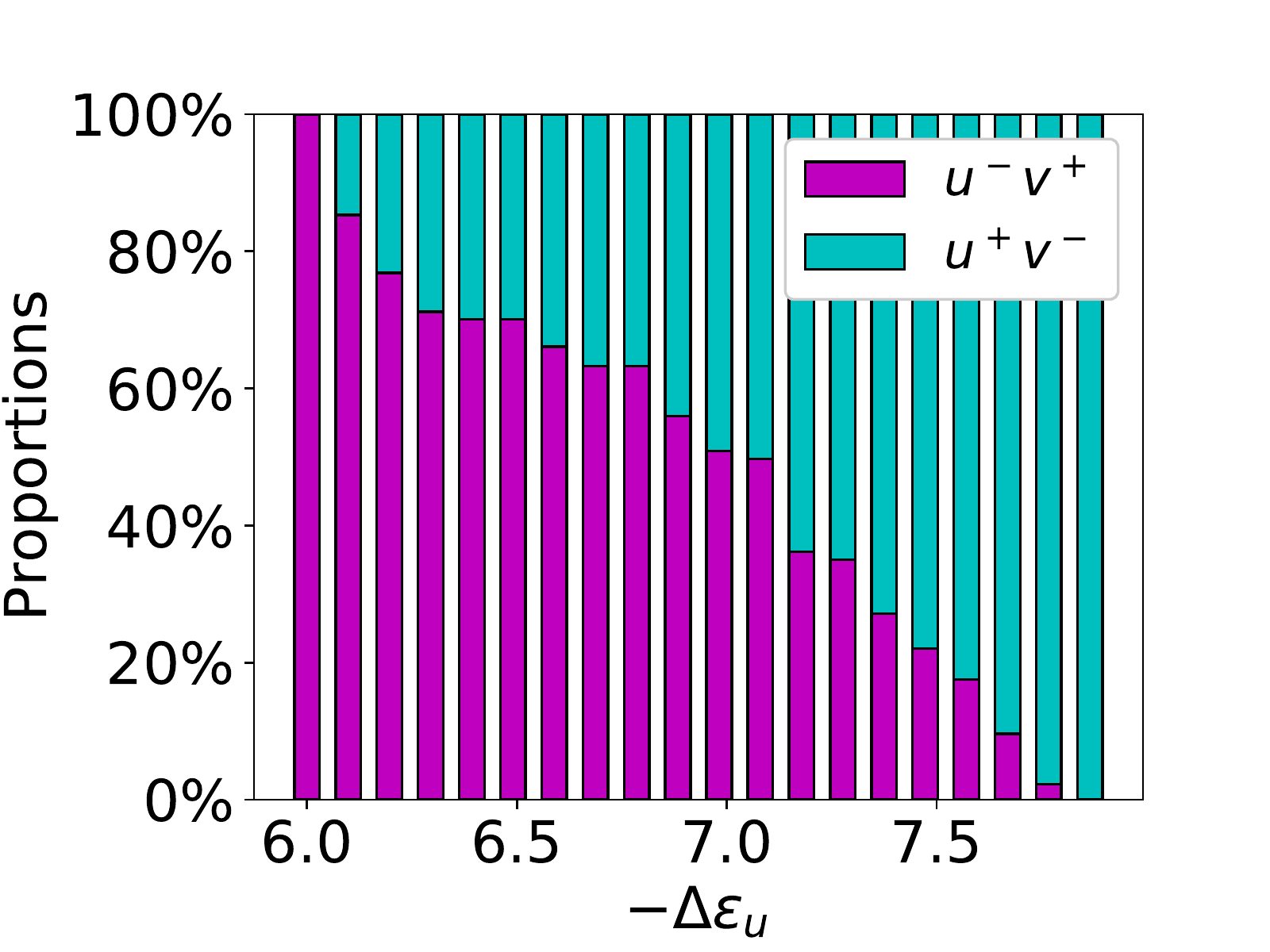}
  \subcaption{$q = 0.1$}
\end{subfigure}
\begin{subfigure}{.32\textwidth}
  \centering
  \includegraphics[width=\linewidth]{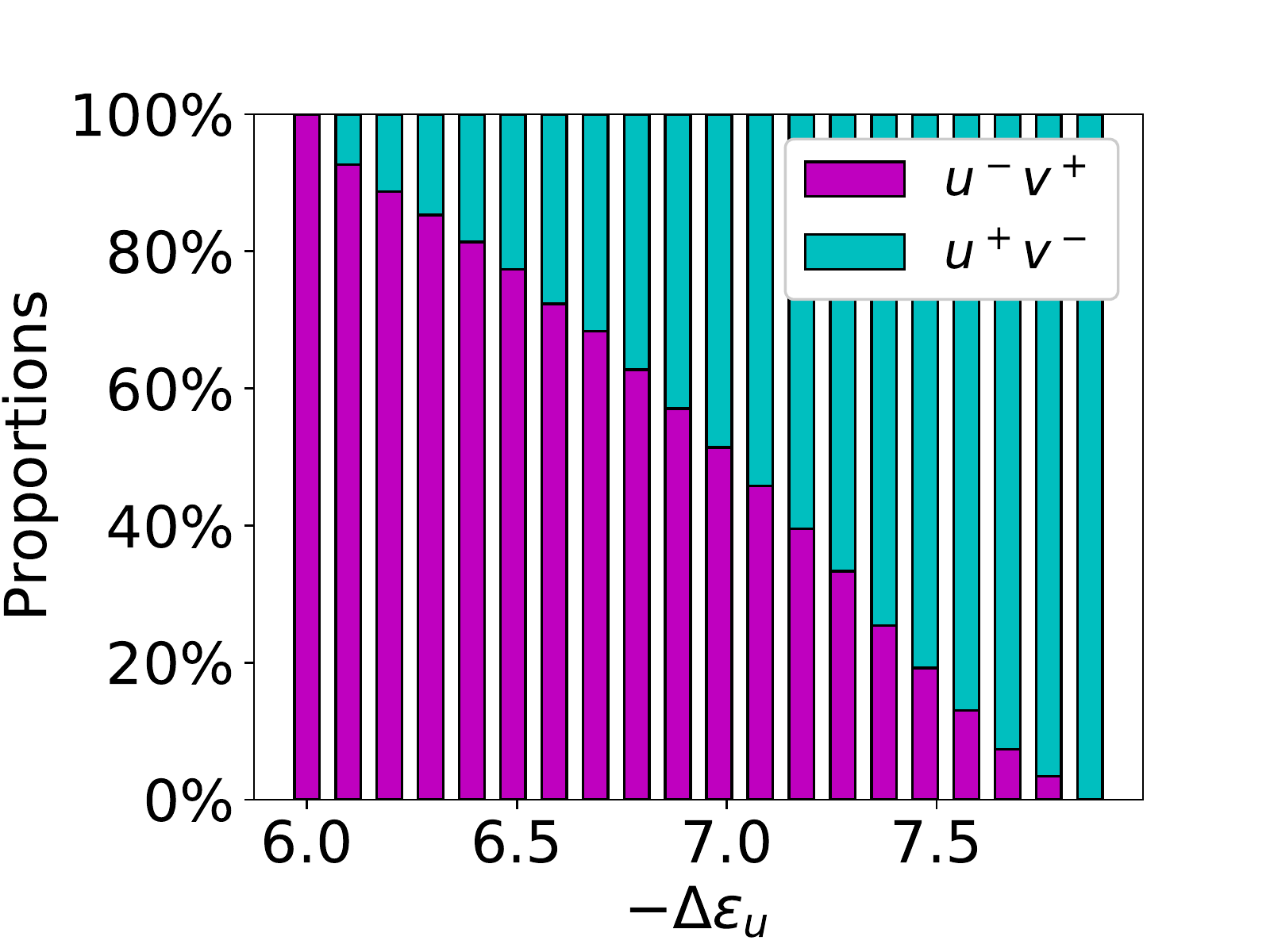}
  \subcaption{$q = 0.5$}
\end{subfigure}
\begin{subfigure}{.32\textwidth}
  \centering
  \includegraphics[width=\linewidth]{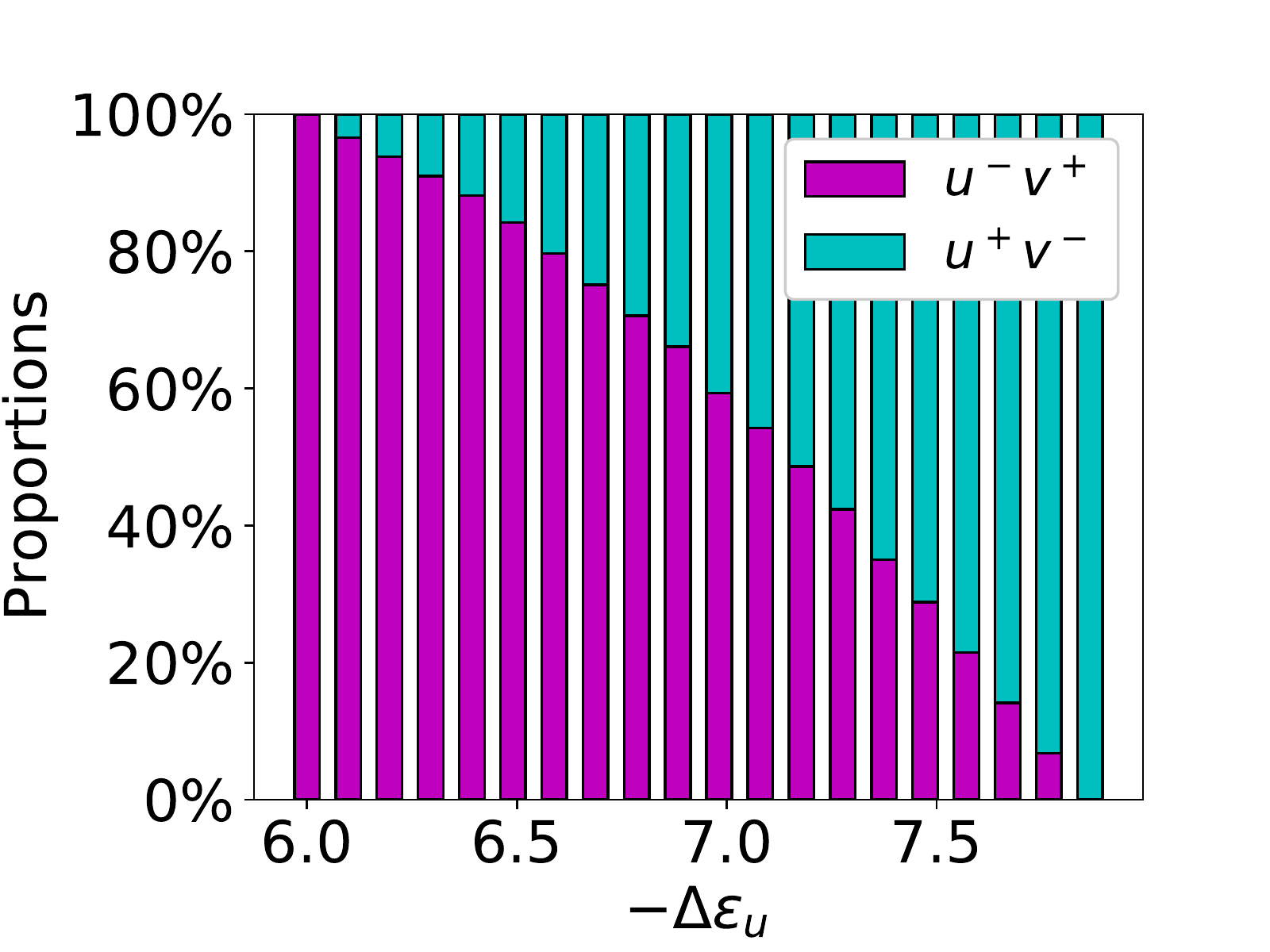}
  \subcaption{$q = 0.9$}
\end{subfigure}
\caption{Simulated cell type proportions with respect to $-\Delta\varepsilon_u$. Simulations were performed by dividing the stability interval for $-\Delta\varepsilon_u$ into $20$ equidistant values. Dispersion parameter $q$ increases from (a) to (c) resulting in different scenarios.}
\label{fig: proportions non-local}
\end{figure}

For different dispersion parameter values $q$, the proportions of $u^-v^+$ show a monotonous decrease with increasing energy difference $-\Delta\varepsilon_u$ (Fig. \ref{fig: proportions non-local}). For low values of $q$, the proportions show some similarities to the local model due to individual larger jumps (Fig. \ref{fig: proportions non-local} (a)). These jumps become less pronounced for medium (Fig. \ref{fig: proportions non-local} (b)) and high dispersions (Fig. \ref{fig: proportions non-local} (c)). Altogether, we have established full control over the cell type proportions.

\section{Discussion}

In this study, we have derived and analyzed a model that allows us to generate cell differentiation patterns based on a system of mutual inhibition of two transcription factors, auto-activation and cell-cell communication. The model was thoroughly analyzed and simulated patterns were characterized. 

\subsection{Derivation of the model from statistical mechanics}

Statistical mechanics has already proven its usefulness in biological model systems like ion channel opening and closing as well as oxygen hemoglobin binding \cite{Garcia2011}. These ideas have further been investigated for transcriptional regulation and were successfully applied for a wide variety of examples \cite{Bintu2005_1, Bintu2005_2}. To our knowledge, cell fate decision models have not been combined with statistical mechanics to date. We derived a specific model based on two mutually inhibiting transcription factors $u$ and $v$ with auto-activation and an external signal inhibiting $u$ and activating $v$. Assuming that auto-activation is the dominant factor in transcriptional regulation, we assume that RNA polymerase binding corresponds to the binding of $u$ and $v$, respectively. Based on this, we were able to derive binding probabilities of RNA polymerase to the respective promoter. A system of ordinary differential equations was generated by combining these probabilities with constant production rates and exponential decay. As long as the auto-activation remains unchanged, minor changes in the GRN such as the removal of either the signal activation or the signal inhibition can still be managed by adjusting the equations accordingly.

\subsection{Analysis of the model allows accurate determination of the stability of heterogeneous steady states}

On the single cell level, we identified that the received signal determines the fate of a cell. There is a critical value of this signal, such that the cell will adopt $u^-v^+$ fate if this value is undercut, or $u^+v^-$ fate if the value is exceeded. This leads to the signal being the relevant factor of the switching behavior in this system. This describes a different point of view compared to systems that utilize differences in initial conditions to generate a cell fate switch \cite{Cherry2000, Huang2007}. At the same time, models that incorporate a signal dependency, have not yet been analyzed in such great detail \cite{Bessonnard2014, Tosenberger2017, Stanoev2021}. Exact expressions for all possible steady states were derived. A stability analysis enabled us to identify parameter values, such that only the states corresponding to $u^+v^-$ and $u^-v^+$ fates are stable. Thus, we were able to limit the system to these two cell fates. On the level of multiple cells, we found analytical expressions of parameter bounds guaranteeing heterogeneous steady states. This means that within these bounds the pattern created by the system will always be a mixture of two different cell types. In conclusion, we have provided the necessary analytical tools to guarantee the generation of heterogeneous patterns of two different cell types.

\subsection{Averaged nearest neighbor signaling leads to checkerboard patterns}

In some biological systems, cell communication is hypothesized to be limited to direct neighbors. An example for this is the lateral inhibition of the Delta Notch signaling pathway in epithelial tissue of \textit{Drosophila}, which has been studied in great detail \cite{Collier1996}. A different example is found in the preimplantation development of the mouse embryo, where transcription factors NANOG and GATA6 decide the fate of cells in the inner cell mass. Computational studies have investigated the effects of activation by an external signal in this biological system \cite{DeMot2016, Bessonnard2014, Tosenberger2017, Stanoev2021}. A great common feature in all of these systems is the formation of checkerboard patterns, i.e. patterns in which cells of one type minimize the number of equal neighbors. Fittingly, we also found this type of pattern in our simulations using an averaged nearest neighbor signaling. We took this one step further and analyzed the possible cell type proportions one can create using this model. An analytical expression for the maximum number of equal cell types in a cell's neighborhood tells us that the cell type proportions are highly linked to the average number of neighboring cells in the system. In our 2D simulations cell type proportions below $30\%$ and above $70\%$ are not possible.

\subsection{Distance-based signaling enables a range of patterns from checkerboard to engulfing}

In addition to the nearest neighbor signal, we investigated the effects of a signal that is capable of being dispersed throughout the tissue. This global cell-cell communication enables a range of patterns. From two cell types in a checkerboard like arrangement to one cell type engulfing the other depending on the signal dispersion. The introduced dispersion parameter $q$ allows us to artificially vary between a signal that only reaches the neighboring cells and a signal that spreads evenly in the tissue. Simulations have shown that for low signal dispersion $u^+v^-$ and $u^-v^+$ cells tend to avoid being adjacent to the same cell type, hence we again recovered the checkerboard pattern. Furthermore, when increasing the signal dispersion, $u^+v^-$ cells accumulate more at the boundary such that overall larger clusters of equal cell types are formed. High signal dispersion leads to an ideal segregation of cells with $u^+v^-$ engulfing $u^-v^+$ cells. Engulfing patterns are often believed to be the result of differential adhesion of two cell types. Indeed, it has already been demonstrated that the minimization of the energy as a function of differential adhesion leads to this type of engulfing \cite{Emily2007}. Not only have we found an alternative way to generate these patterns, but at the same time we were able to unify the formation of both checkerboard and engulfing patterns under the notion of differently dispersing signals.

\subsection{Conclusion}

We have provided a new model to describe transcriptional regulation for a system of mutually exclusive transcription factors. Furthermore, the model was analyzed in great detail with respect to parameters and stability. The model was extended by signaling mechanisms describing the cell-cell communication. The local and global signaling obey a simple mathematical rule depending on the number of cells it has to travel across in order to reach its destination. A detailed description of the signaling transport mechanism, possibly including diffusion and advection mechanisms, provides room for further research. Additionally, signal production and uptake of cells play a crucial role in how effective different means of signal transport might be. Another perspective can be achieved by incorporating cell growth and cell division into the model and analyzing their effect on the resulting patterns. With this in mind, our study paves the way for numerous subsequent studies regarding signal-based pattern formation in developmental systems.

\section*{Acknowledgements}

We thank Nicholas A.M. Monk for his detailed feedback on the analytical part of this study.

\bibliography{mybib}

\end{document}